\documentclass[]{aa}
\usepackage{graphicx}
\usepackage{txfonts}
\usepackage{lscape}
\usepackage{rotating}

\begin{document}

\title{Modeling the spectral energy distribution of ULIRGs I: the radio spectra}

\author{M.S. Clemens
       \inst{1}
       \and
       O. Vega\inst{1,2}
       \and
       A. Bressan\inst{1,2,3}
       \and
       G.L. Granato\inst{1,3}
       \and
       L. Silva\inst{4}
       \and
       P. Panuzzo\inst{1,5}}

\offprints{M.S. Clemens, \email{marcel.clemens@oapd.inaf.it}}

\institute{INAF, Osservatorio Astronomico di Padova, Vicolo dell'Osservatorio, 5,
           35122 Padova, Italy
           \and
           INAOE, Luis Enrique Erro 1, 72840 Tonantzintla, Puebla, Mexico
           \and
            SISSA, Strada Costiera, I-34131 Trieste, Italy
           \and
           INAF, Osservatorio Astronomico di Trieste, Via Tiepolo 11, I-34131 Trieste, Italy
           \and
           Laboratoire AIM, CEA/DSM - CNRS - Universit\'{e} Paris Diderot, DAPNIA/Service 
           d'Astrophysique, B\^{a}t. 709, CEA-Saclay, F-91191 Gif-sur- Yvette C\'{e}dex, France}

\date{Received ????; accepted ????}

\abstract
{}
{We aim to constrain new starburst/AGN models of IRAS bright galaxies via their spectral energy distribution from the near-infrared to the radio. To this end, we determine the radio spectra for a sample of 31 luminous and ultraluminous IRAS galaxies (LIRGs/ULIRGs).}
{We present here new high frequency VLA observations at 22.5~GHz and 8.4~GHz and also derive fluxes at other radio frequencies from archival data. Together with radio data from the literature, we construct the radio spectrum for each source. In the selection of data we have made every effort to ensure that these fluxes neither include contributions from nearby objects, nor underestimate the flux due to high interferometer resolution.}
{From our sample of well-determined radio spectra we find that very few have a straight power-law slope. Although some sources show a flattening of the radio spectral slope at high frequencies, the average spectrum shows a \emph{steepening} of the radio spectrum from 1.4 to 22.5~GHz. This is unexpected, because in sources with high rates of star formation, we expect flat spectrum, free-free emission to make a significant contribution to the radio flux at higher radio frequencies. Despite this trend, the radio spectral indices between 8.4 and 22.5~GHz are flatter for sources with higher values of the far-infrared (FIR)-radio flux density ratio, $q$, when this is calculated at 8.4~GHz. Therefore, sources that are deficient in radio emission relative to FIR emission (presumably younger sources) have a larger thermal component to their radio emission. However, we find no correlation between the radio spectral index between 1.4 and 4.8~GHz and $q$ at 8.4~GHz. Because the low frequency spectral index is affected by free-free absorption, and this is a function of source size for a given mass of ionized gas, this is evidence that the ionized gas in ULIRGs shows a range of densities.}
{The youngest LIRGs and ULIRGs are characterized by flatter average radio spectral indices from 1.4 to 22.5~GHz, and by a larger contribution to their high frequency, radio spectra from free-free emission. However, the youngest sources are not those that have the greatest free-free absorption at low radio frequencies. The sources in which the effects of free-free absorption are strongest are instead the most compact sources. Although these have the warmest FIR colours, they are not necessarily the youngest sources.}


\maketitle

\section{Introduction}
\label{sec:intro}
The nature of the power source for ULIRGs remains a much debated issue. 
The presence of either an AGN or a starburst in a given source is not 
evidence that one or the other is the principal power source for the 
infrared luminosity. The observation that ULIRGs fall on the same 
FIR-radio correlation as star-forming galaxies (e.g. Sopp, Alexander
 \& Riley, 1990) was taken as evidence that these sources are powered 
predominantly by star formation. Analogous correlations exist between 
the Br$\gamma$ luminosity and far-infrared luminosity (Goldader et al. 1997) 
and dense molecular gas mass and far-infrared luminosity (Gao \& Solomon, 2004).
However, evidence that AGN activity is not independent of starburst activity 
have made such conclusions less secure. Farrah et al. (2003) show, in fact, that AGN 
and starburst luminosities are correlated over a wide range of IR luminosities. 
As ULIRGs may commonly host both an AGN and a starburst, their relative 
contributions need to be quantified.

Apart from the rarity of ULIRGs (there is no example closer than Arp 220 at 72 Mpc) their
study is made difficult by the large and uncertain extinctions toward their centres.
In many cases the extinction is not even negligible in the mid-infrared; it is likely
that $\rm H\alpha$ emission detected from a ULIRG does not originate from the deeply
embedded regions where the FIR luminosity is generated. The less extinguished, outer
regions of a ULIRG may host a certain rate of star formation that causes a certain
amount of $\rm H\alpha$ emission but this could be quite independent of whether it
hosts a central starburst or an AGN. Murphy et al. (2001) find very few broad near-infrared
lines in ULIRGs that would be the signature of an AGN, but note that very high extinctions
could exclude their detection even at $2\;\rm \mu m$. Flores et al. (2004) recently showed
that even when corrected by $\rm H\gamma$ data, the $\rm H\alpha$ luminosity can underestimate
the star formation rate by a factor of 2 for ULIRGs. Both Goldader et al. (1995)
and Vald\'es et al. (2005) find that near-infrared recombination lines (Pa$\alpha$
and Br$\gamma$) are under-luminous in ULIRGs compared with what would be expected
from a starburst of similar bolometric luminosity. Their results indicate that
even near-infrared observations may not penetrate the most obscured regions in these sources.
This picture is consistent with the conclusions of Poggianti et al. (2001)
that the extinction in the centre of ULIRGs
is a function of the age of the star formation episode,
with younger stars being more heavily extinguished, as introduced by 
Silva et al. (1998) to describe star forming regions. 
The lack of correlation
between super star clusters and HII regions found in luminous infrared
galaxies by Alonso-Herrero et al. (2002) also supports a scenario of age-dependent 
extinction. In order to estimate the star formation rate in these sources tracers that 
do not suffer extinction are preferred.

Perhaps the most direct way to determine the power source in ULIRGs is via 
radio recombination lines that trace the ionized gas without suffering extinction.
Observations of several lines between 1.4 and 207~GHz of Arp~220 have been used by 
Anantharamaiah et al. (2000) to show that this source is powered by star formation.
Unfortunately, very few ULIRGs are accessible observationally to these kind of 
observations with existing instruments.

One approach which has been used to distinguish between AGN and starburst
power sources in ULIRGs has been to search for very compact radio continuum
emission towards the nucleus or nuclei. Interestingly, radio sources of similar 
\emph{physical} sizes have been identified as both AGN and compact starbursts, 
depending on the resolution of the observations. Nagar et al. (2003) use 15~GHz 
radio continuum data with a resolution of 150 mas to investigate the nature of 
83 ULIRGs. One argument used by these authors to conclude that most ULIRGs are 
AGN powered is the compactness of the radio sources detected. The resolution of 
their data corresponds to 420 pc at the median redshift of their sample. However, 
the supernovae (SNe) detected in the NW nucleus of Arp~220 (Smith et al., 1998; Lonsdale et 
al. 2006) are within a region $0\, \farcs 2 \times 0\, \farcs 4$ ($75 \times 150 \;\rm pc$). 
Both of these studies found that no AGN is necessary to explain the IR luminosity 
of Arp~220. As long as the brightness temperature does not exceed $10^{6} \;\rm K$ 
the compactness of radio nuclei alone does not support an AGN hypothesis.

Rather than look at the radio morphology, we examine the spectrum of the radio emission.
In star-forming galaxies the radio spectrum is made-up of two components: a non-thermal, synchrotron component and a thermal, `Bremsstrahlung' component, often referred to as `free-free' emission. At GHz frequencies, the synchrotron spectrum, $S_{\nu} \propto \nu^{\alpha}$, has a typical power-law slope, $\alpha \sim -0.8$, while the free-free emission has a much flatter slope, $\alpha = -0.1$. Free-free emission thus makes a larger contribution at higher radio frequencies, so that the radio spectrum flattens at high frequencies. Half of the radio flux is expected to be of thermal origin at $\sim 20\;\rm GHz$ (Condon, 1992). Towards lower radio frequencies (below 1~GHz in non-starburst galaxies) the free-free optical depth becomes large and the (synchrotron-dominated) radio spectrum shows a sharp decline due to free-free absorption (see Fig.4 of Condon, 1992). In sources with very intense star-formation, free-free absorption may be expected to occur at higher frequencies than in quiescent objects; the 1.4~GHz fluxes from such objects are probably affected.

In principle, a measure of the thermal radio flux from a ULIRG would be an
excellent measure of the star formation rate, because, being emitted by the same
gas from which recombination lines originate, it traces gas
ionized by young stars. Absorption of ionizing photons by dust may complicate 
this picture slightly (Vald\'{e}s et al, 2005) but the main difficulty is in estimating 
the thermal fraction. Although
the fractional contribution from thermal emission to the total radio flux increases
with frequency, the radio emission from ULIRGs is dominated by synchrotron
radiation even at frequencies above 15~GHz. As we will see later, the radio spectral
indices around 15~GHz are normally much steeper than the -0.1 expected from a purely
thermal spectrum. The motivation for the 22~GHz observations described here was to better
constrain the thermal radio flux.

Because the thermal radio flux provides an estimate of the present
star-formation rate, it is important in fitting starburst/AGN models to
the spectral energy distribution (SED) from the near-infrared to the
radio. Bressan, Silva \& Granato (2002) showed that deviations from
the FIR/radio correlation could be used to derive the evolutionary
status of a compact starburst. Such deviations are expected in bursts
of short duration because at early times (a few $\times 10^6\;\rm yr$)
even the most massive stars formed in the burst will not have ended
their lives as supernovae and no excess synchrotron emission should
result from the starburst activity. Such young sources then, should
have a FIR/radio ratio above the mean FIR-radio correlation and should
have flatter radio slopes due to the greater ratio of
thermal/synchrotron emission. In Prouton et al. (2004) we illustrated
the power of this technique with special emphasis on high frequency
radio data. A second paper, Vega et al. (2007), hereafter paper 2,
will be concerned with the model fits to the SEDs from the
near-infrared to the radio.

Here we present new radio data at 22.5 and 8.4~GHz, as well as archival radio data which has been re-reduced, and radio fluxes from the literature. The result is a set of 31, well-sampled  radio spectra for infrared-luminous galaxies. In \S~\ref{sec:sample} we describe the sample selection and in \S~\ref{sec:obs} the data acquisition and reduction. In \S~\ref{sec:results} we describe the radio spectra in detail, and in \S~\ref{sec:discussion}, the implications of these spectra when compared to far-infrared fluxes and the FIR-radio flux ratio, $q$. Conclusions are reached in \S~\ref{sec:conclusions}. In an appendix we comment on individual objects and present the radio spectra in tabular form.

\section{Sample selection}
\label{sec:sample}
We have based our sample on that of Condon et al. (1991a) who made 8.4~GHz
observations of the 40 ULIRGs brighter than $5.25\;\rm Jy$ at $60\;\rm \mu m$ 
in the IRAS Bright Galaxy Sample.
In Prouton et al. (2004) we observed 7 of the ULIRGs observed by Condon et al. 
but restricted our attention to objects with a ratio of FIR to $1.4\;\rm GHz$ 
luminosity $q \geq 2.5 $, in an attempt to preferentially select young starbursts. 
Here we have relaxed this restriction on $q$ so as to maximize the sample size 
and include more diverse objects, and have 
observed a further 18 sources at 22.5~GHz. The remaining sources were either 
too weak for observation at 22~GHz or were too confused with nearby sources to 
construct a reliable SED. We also re-observed 7 sources at 8.4~GHz for which the 
high resolution observations of Condon et al. (1991a) could not provide reliable 
integrated fluxes. Our final sample is therefore not a far-infrared flux limited 
sample like that of Condon et al., but simply contains all those brighter than 
$5.25\;\rm Jy$ at $60\;\rm \mu m$ for which an integrated 22.5~GHz flux could be 
reliably obtained.  

In addition to the new high frequency observations, we have obtained,
either from the literature or from reduction of archival VLA data,
radio fluxes at as many frequencies as possible. In total, of the 40
ULIRGs of the sample of Condon et al., 31 have radio data at 3
or more frequencies and 29 of these have fluxes at 22.5~GHz.   

All our sample galaxies are at a redshift $z<0.1$.

\section{Observations and data reduction}
\label{sec:obs}
\subsection{New 22.5 and 8.4~GHz data}
22.5~GHz observations were carried out on 2004, August 13 and 14 with
the VLA in D-configuration. Total integration times for the sources were
between 16 and 24 minutes, with scans on-source being interleaved between scans of
phase calibrators every 8 minutes. Before each new source was observed, a pointing
scan was carried out on the phase calibrator at 8.4~GHz in order to determine a
pointing correction. 3C~286 was observed as an absolute flux calibrator on 13 August 
and 3C~48 on 15 August.

The weather conditions on August 13 were not ideal for high frequency radio observations,
data were taken in the presence of thunder storms. As water vapour affects the measured
flux for a given source at 22~GHz, and the flux calibrator may have been observed through a
different cloud density than some of the phase calibrators, the bootstrapping of the phase
calibrator fluxes to those of the flux calibrators is not guaranteed to work. After flux
calibration based on the {\sc clean} component models for 3C~286 and 3C~48 provided by NRAO the
phase calibrator fluxes were within 20\% of those given by the VLA calibrator database.
Only in the case of 1310+323 (the phase calibrator for UGC~8387) was there a larger
discrepancy (a factor of 2 less than given in the database). 

\begin{figure*}
\centering
\includegraphics[angle=-90, scale = 0.8]{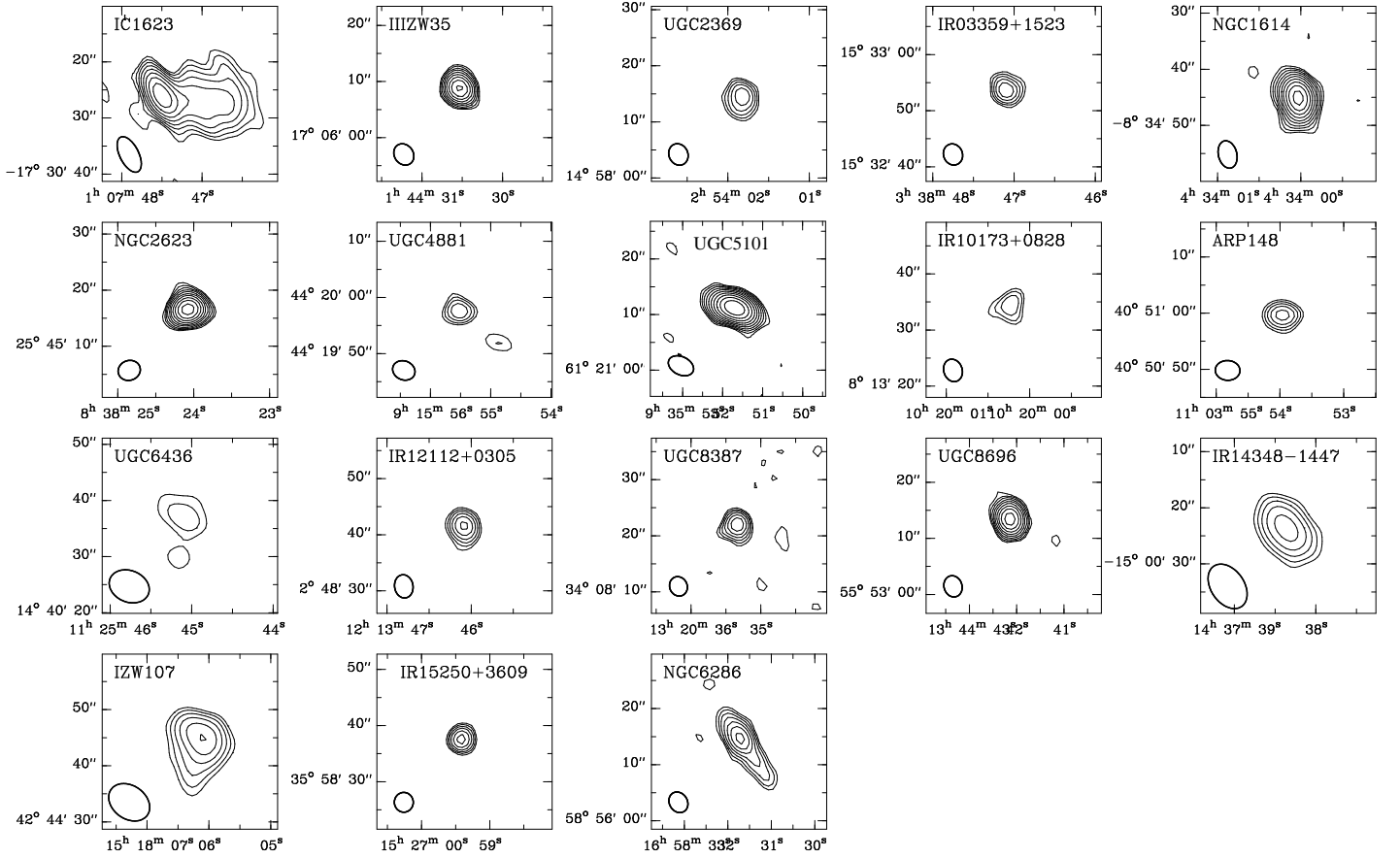}
\caption{22.5~GHz images of the newly observed ULIRGs in our sample. Contour levels begin at $\pm 0.5 \;\rm mJy\, beam^{-1}$ and are spaced by a factor of $2^{1/2}$. The synthesized beam is shown in the  bottom left of each panel. }
\label{Fig:sources}
\end{figure*}

Maps were made using 'natural' weighting in the uv-plane and in the case of UGC~6436,
IZW~107 and IRAS~14348-144 a uv-taper of $30\;\rm k\lambda$ was applied to avoid the
possibility of missing flux due to extended but weak structure. The final maps, shown in 
Fig.~\ref{Fig:sources}, had a typical resolution of 4$^{\prime\prime}$ (6$^{\prime\prime}$ 
where tapered) and after {\sc clean}ing had a typical rms noise level of $0.12 \;\rm mJy/beam$. 
Fluxes were measured by direct integration on the maps. These fluxes (reported in 
table~\ref{tab:cal}) have errors which include the rms noise and the flux calibration errors
described above.

\begin{figure*}
\centerline{
\hskip10mm
\includegraphics[scale=0.27]{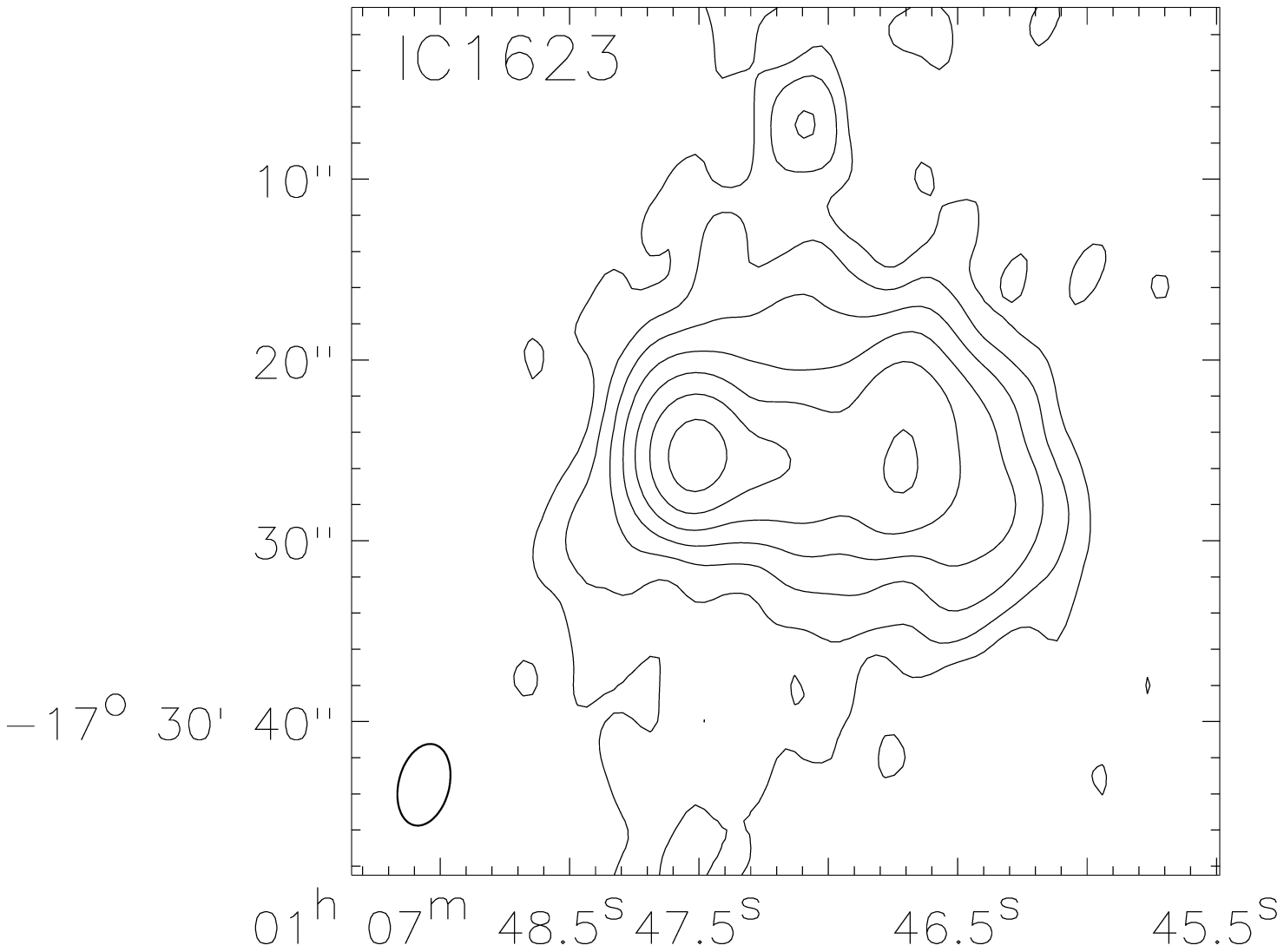}
\hskip-8mm
\includegraphics[scale=0.27]{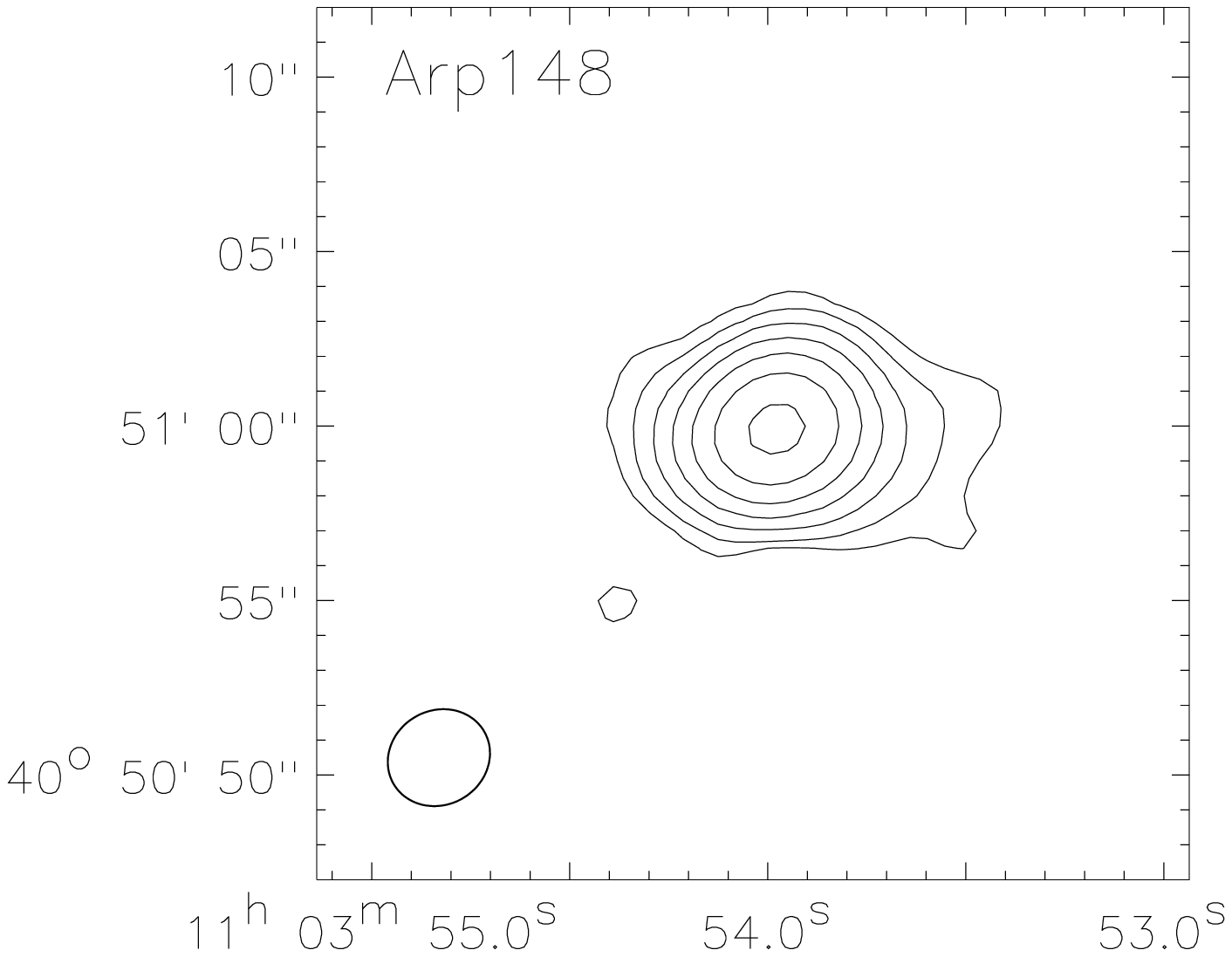}
\hskip-8mm
\includegraphics[scale=0.27]{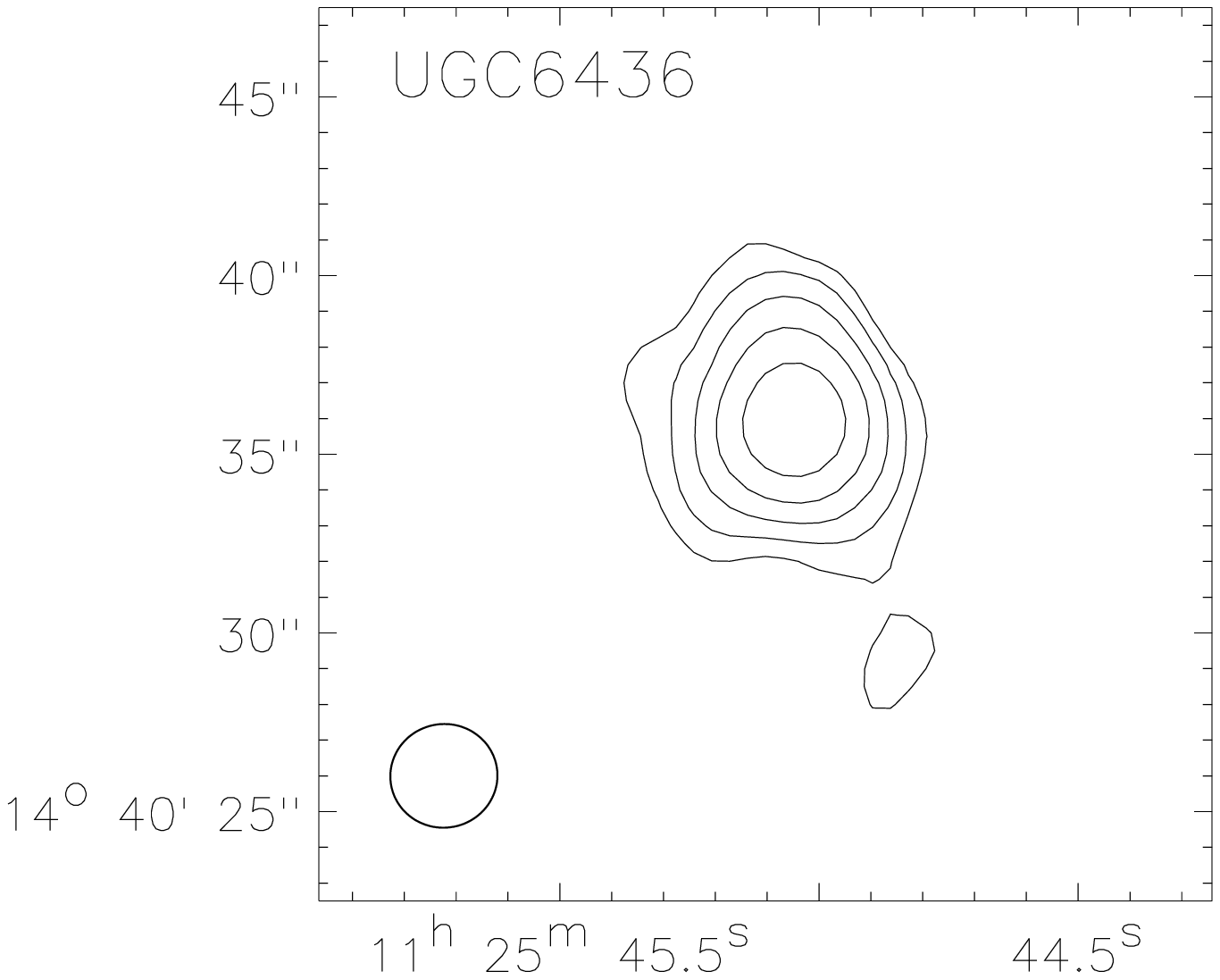}
\hskip-8mm
\includegraphics[scale=0.27]{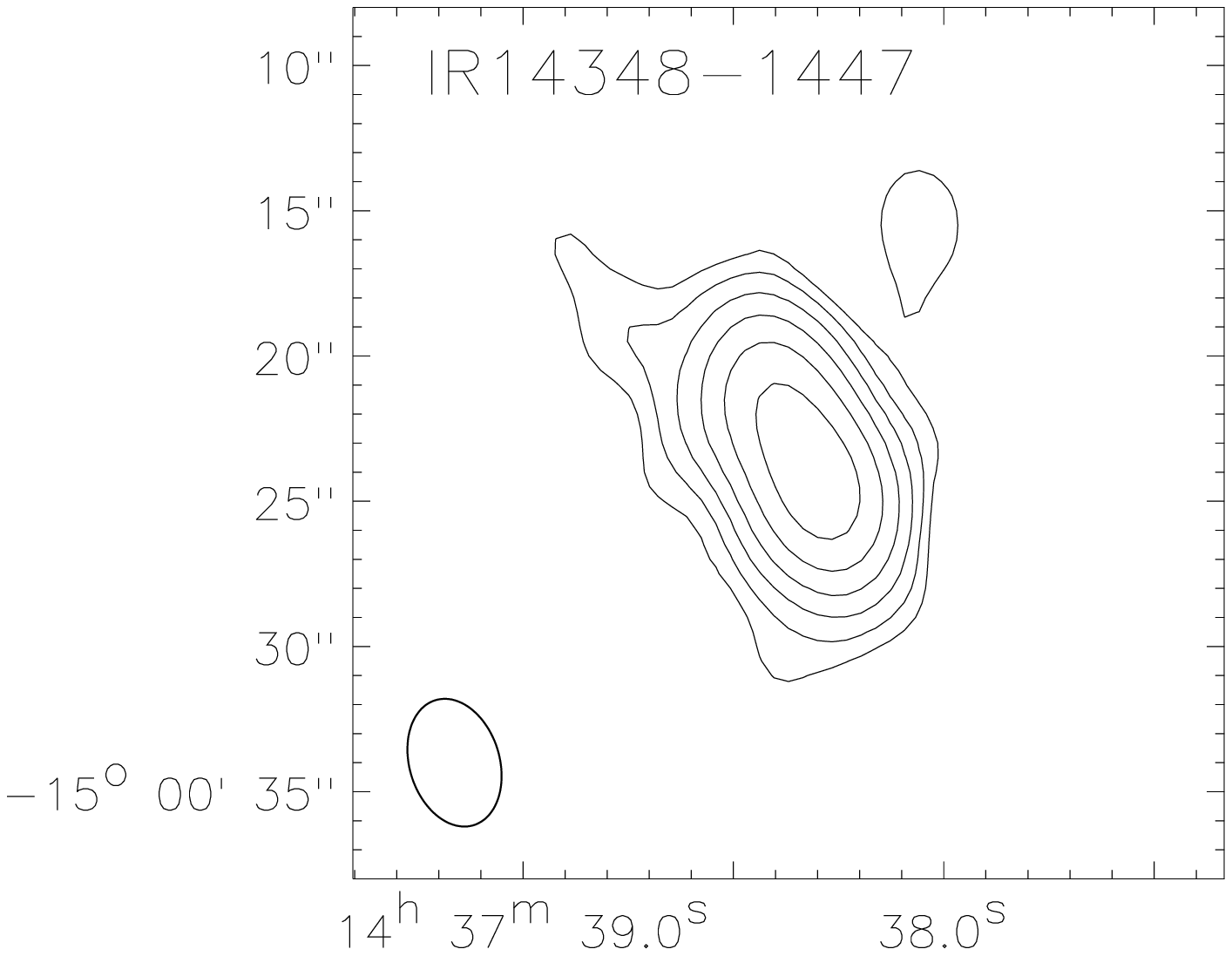}
}
\centerline{
\hskip10mm
\includegraphics[scale=0.27]{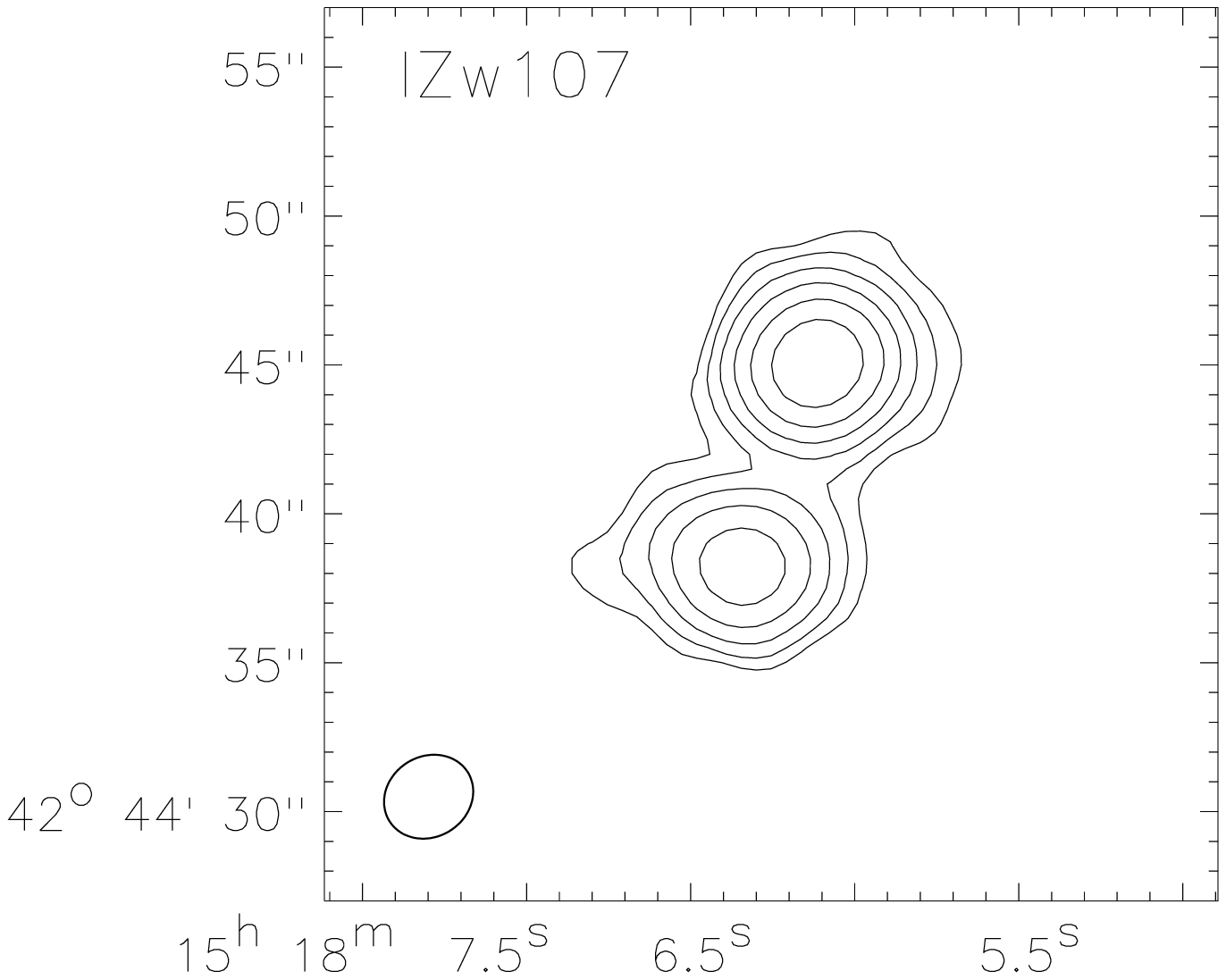}
\hskip-8mm
\includegraphics[scale=0.27]{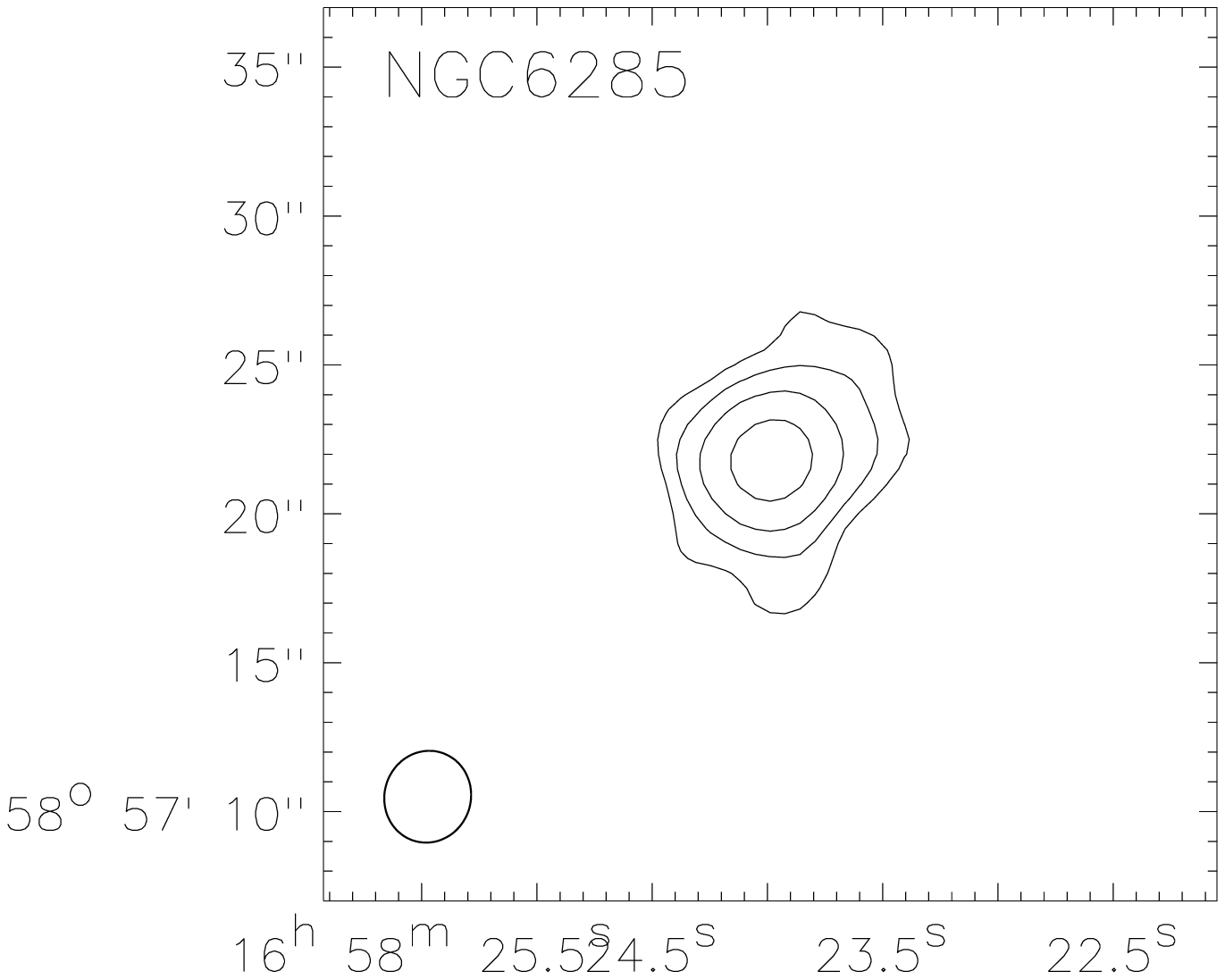}
\hskip-8mm
\includegraphics[scale=0.27]{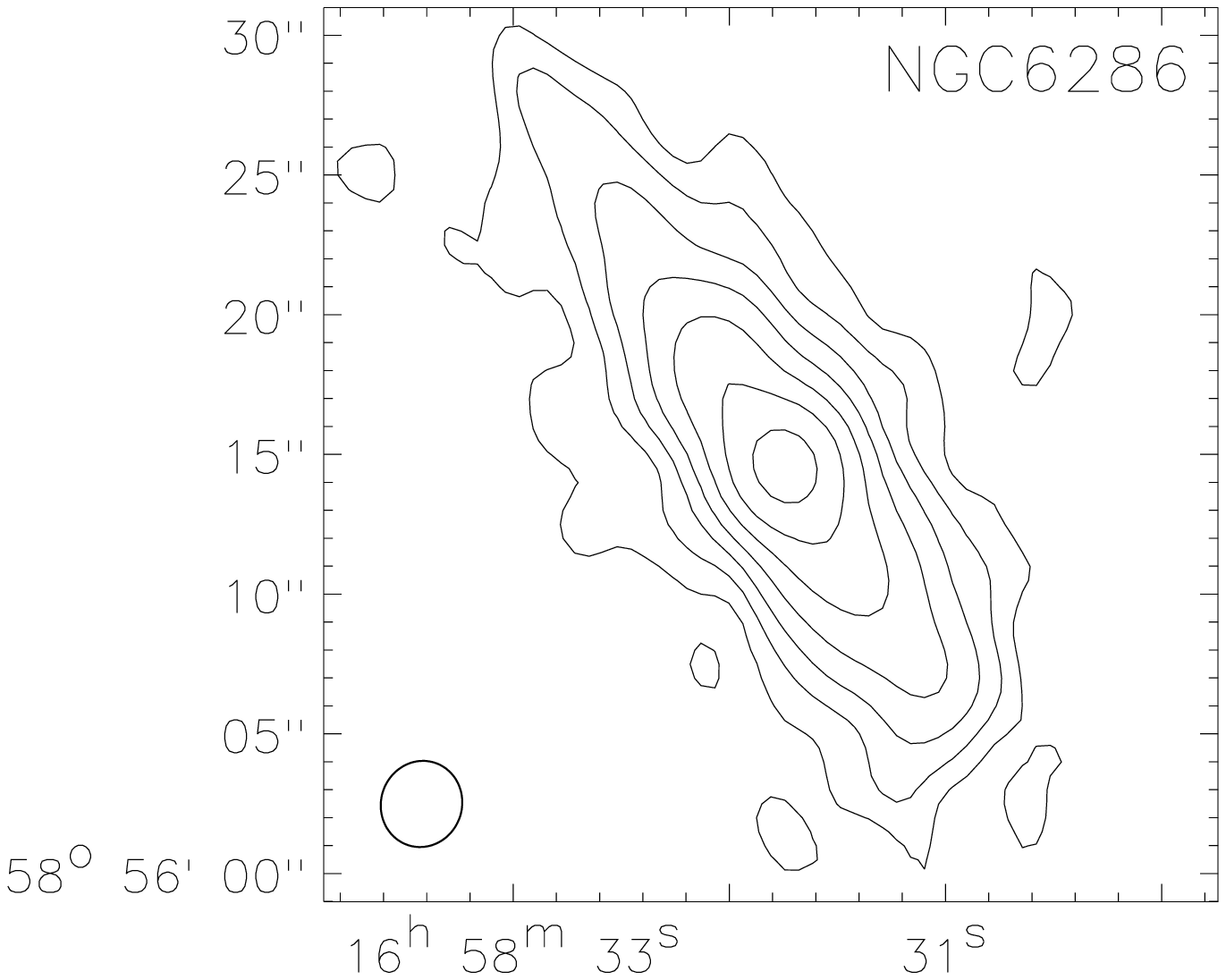}
\hskip-8mm
\includegraphics[scale=0.27]{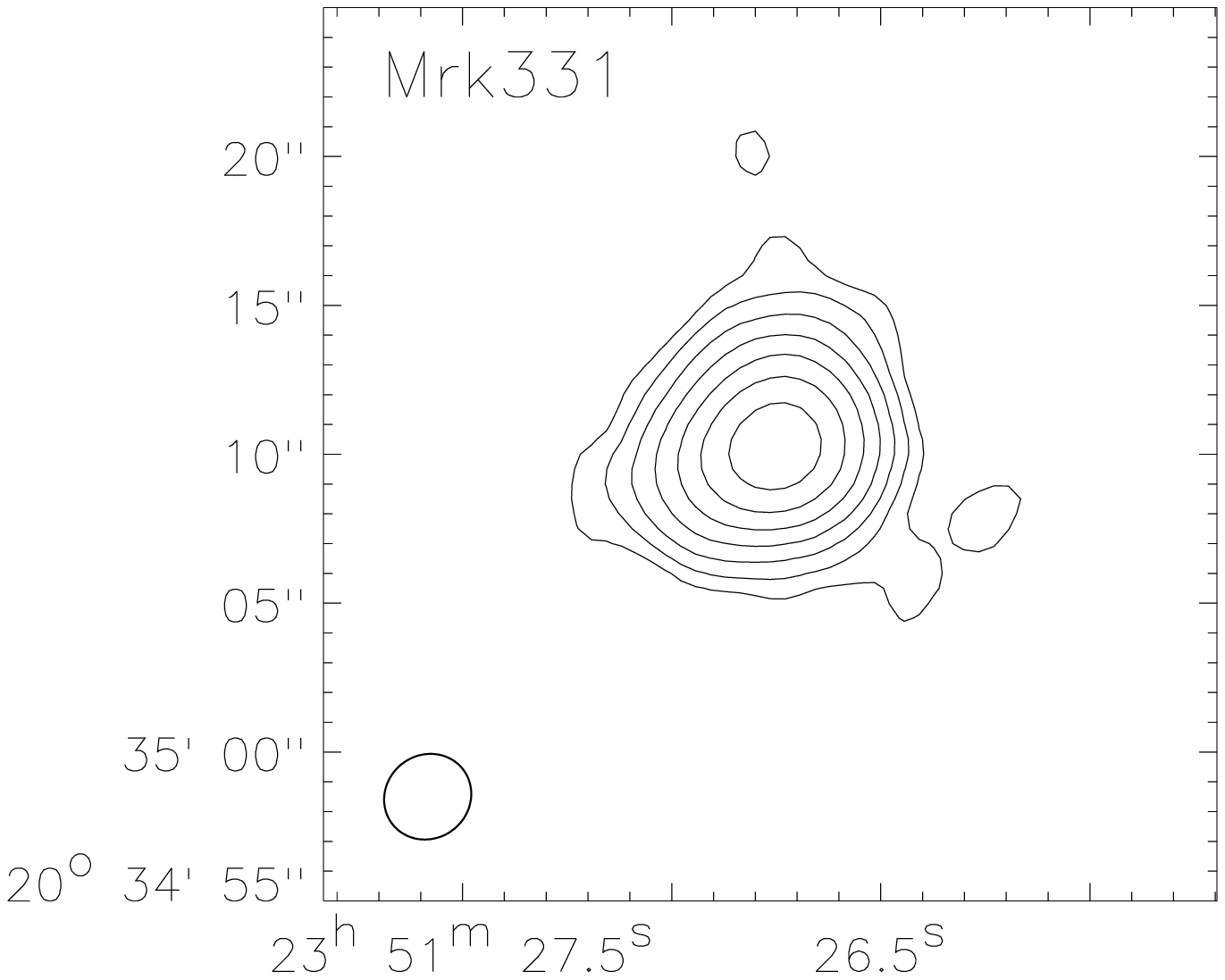}
}
\caption{Contour maps for the new 8.4~GHz observations. The synthesized beam is shown in the  bottom left of each panel. NGC~6285 is the companion of NGC~6286 and its radio map is included only for completeness. }
\label{fig:844Ghz}
\end{figure*}

8.44~GHz observations were made on 2005, July 12, 20 and 27 with the C-configuration of the
VLA. Integration times of 11 to 14 minutes were used and interleaved with observations of nearby
phase calibrators. 3C~286 and 3C~48 were used as absolute flux calibrators and, as for the 22~GHz 
data, calibration made use of {\sc clean} component models for these sources. Derived fluxes for
 all phase calibrators agreed with those in the VLA calibrator database to better than 5\%. All 
sources were also self-calibrated to correct for phase only.  Maps were made by natural weighting 
the uv-data and {\sc clean}ed in the standard way. Resolution and rms in the resulting maps were 
typically $3^{\prime\prime}$ and $0.03\; \rm mJy\,beam^{-1}$. The radio maps for these sources are 
presented in Fig.~\ref{fig:844Ghz} and the corresponding integrated fluxes in table~\ref{tab:8GHz}.

\subsection{Archival VLA data}
\begin{figure*}
\centerline{
\includegraphics[scale=0.38]{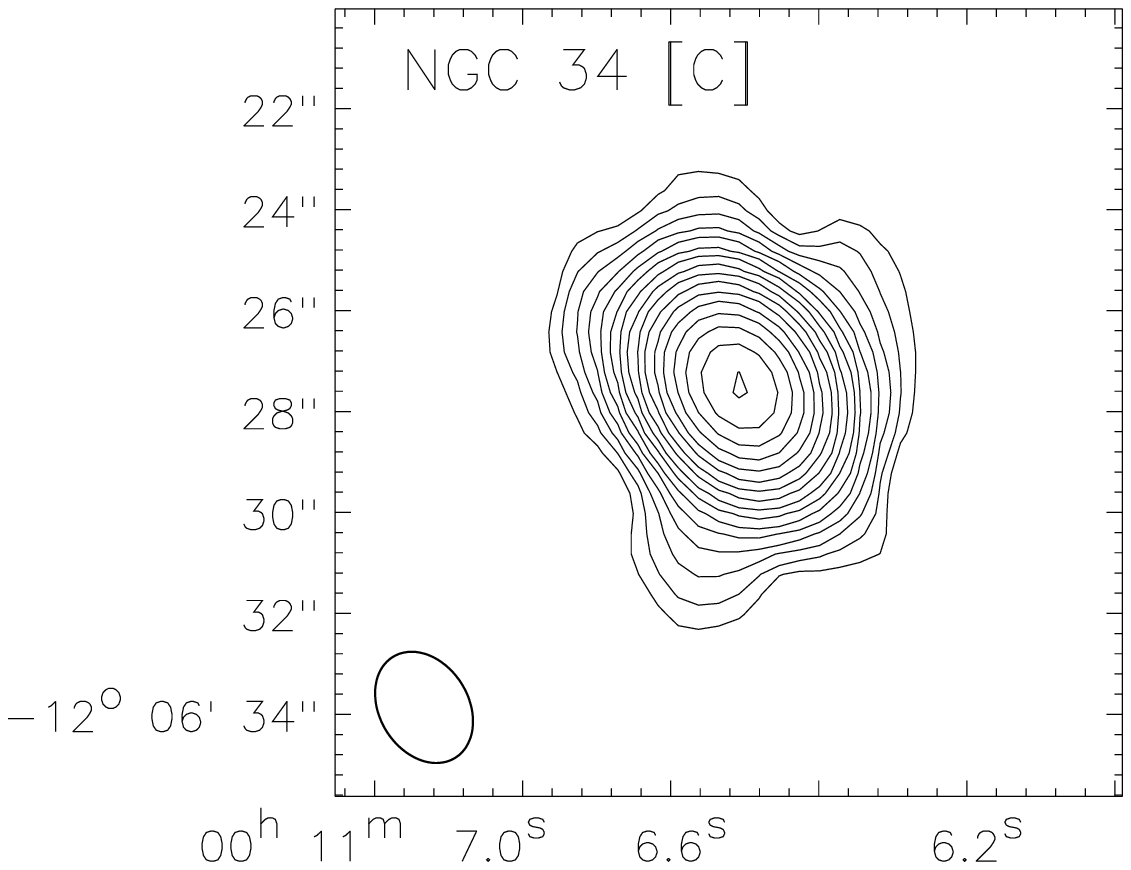}
\includegraphics[scale=0.38]{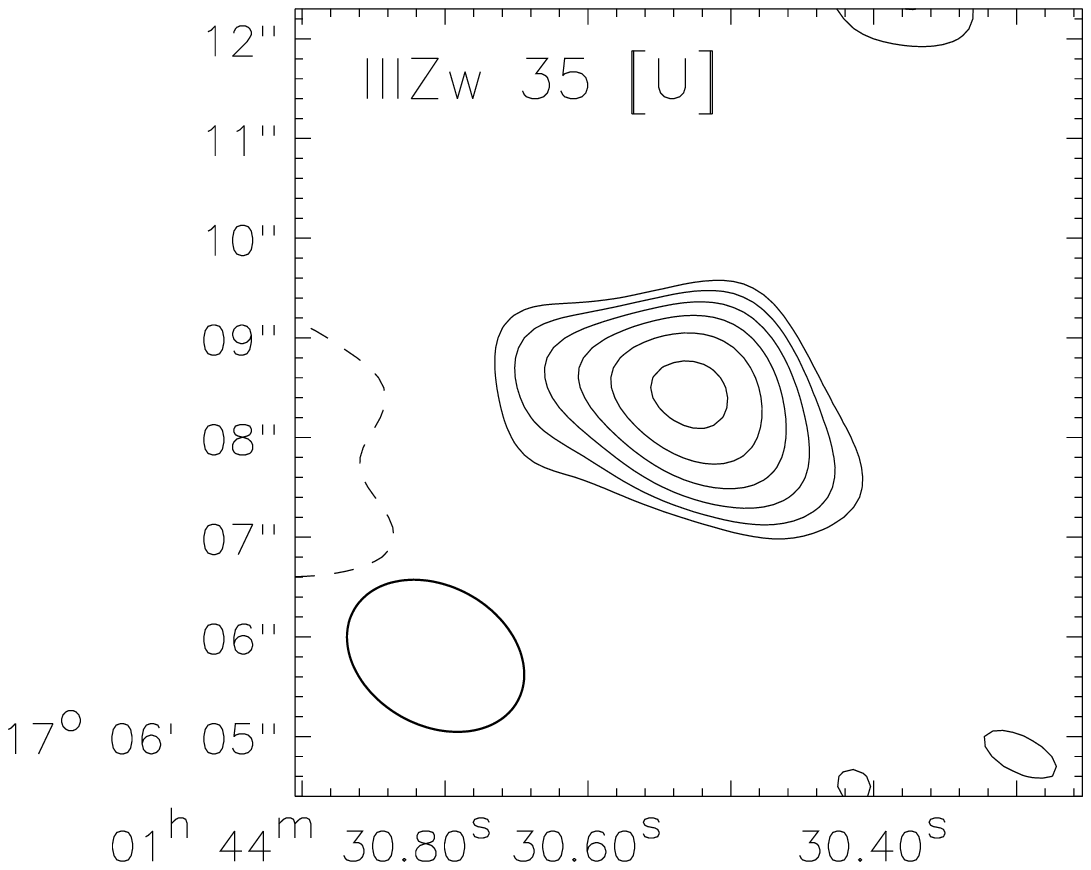}
\includegraphics[scale=0.38]{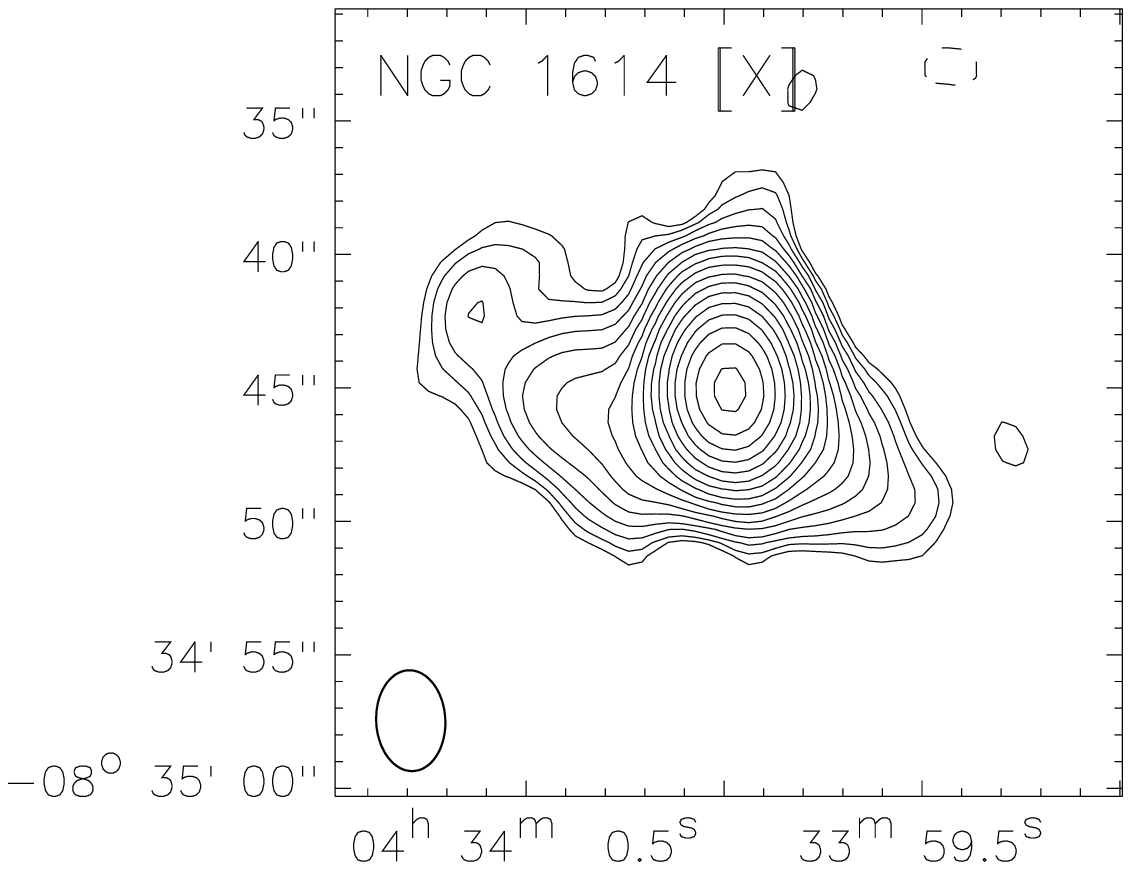}
\includegraphics[scale=0.38]{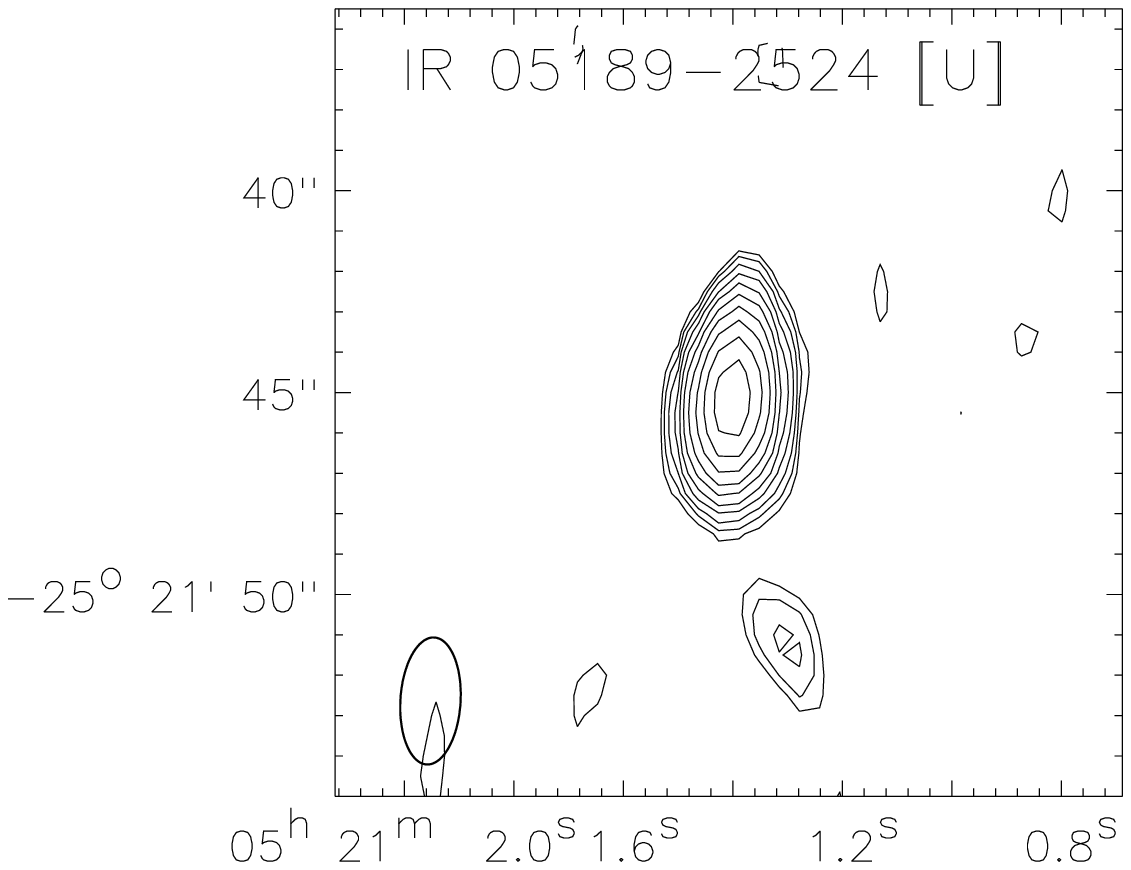}
}
\centerline{
\includegraphics[scale=0.38]{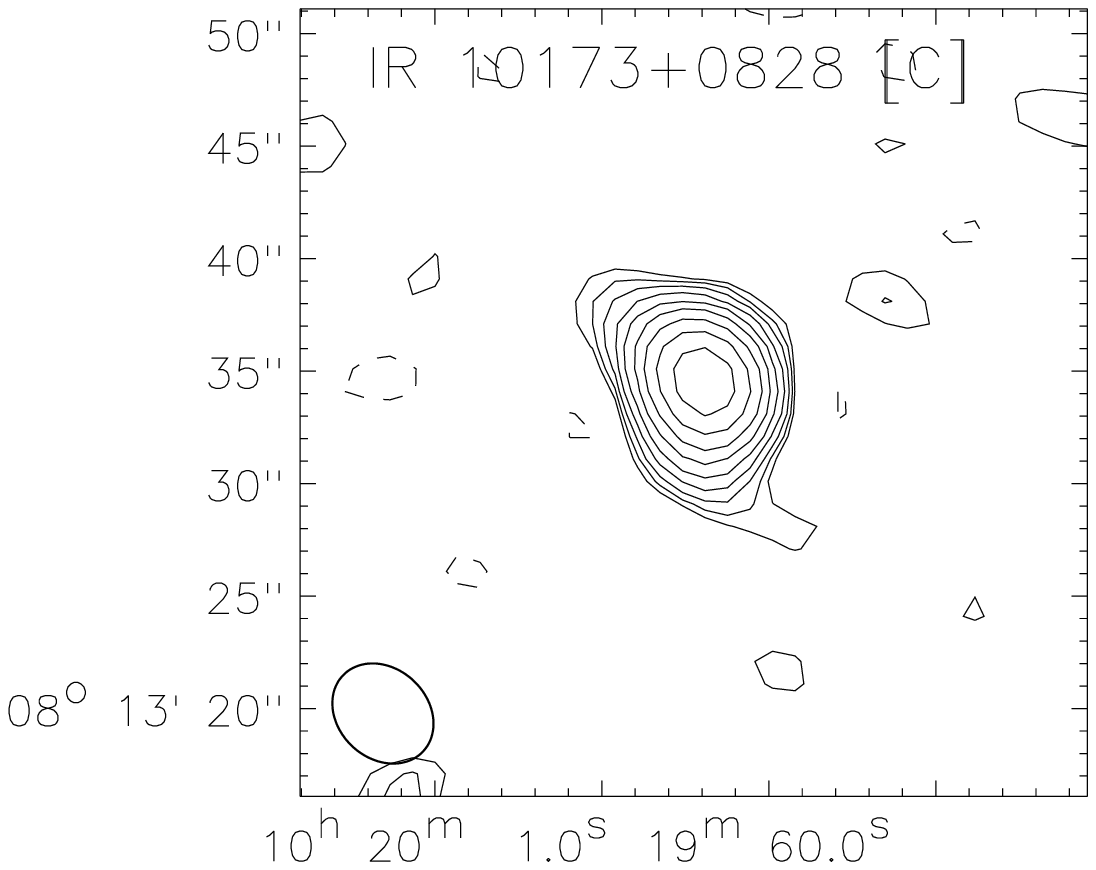}
\includegraphics[scale=0.38]{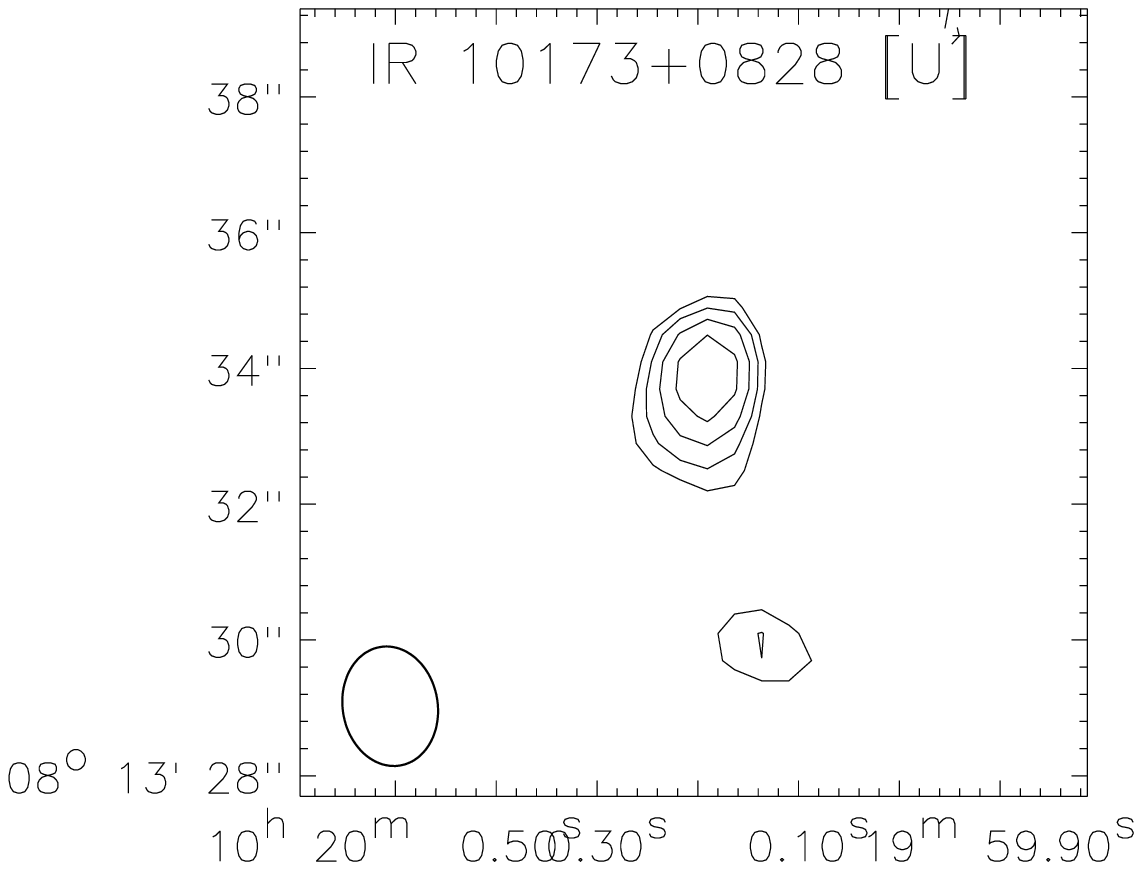}
\includegraphics[scale=0.38]{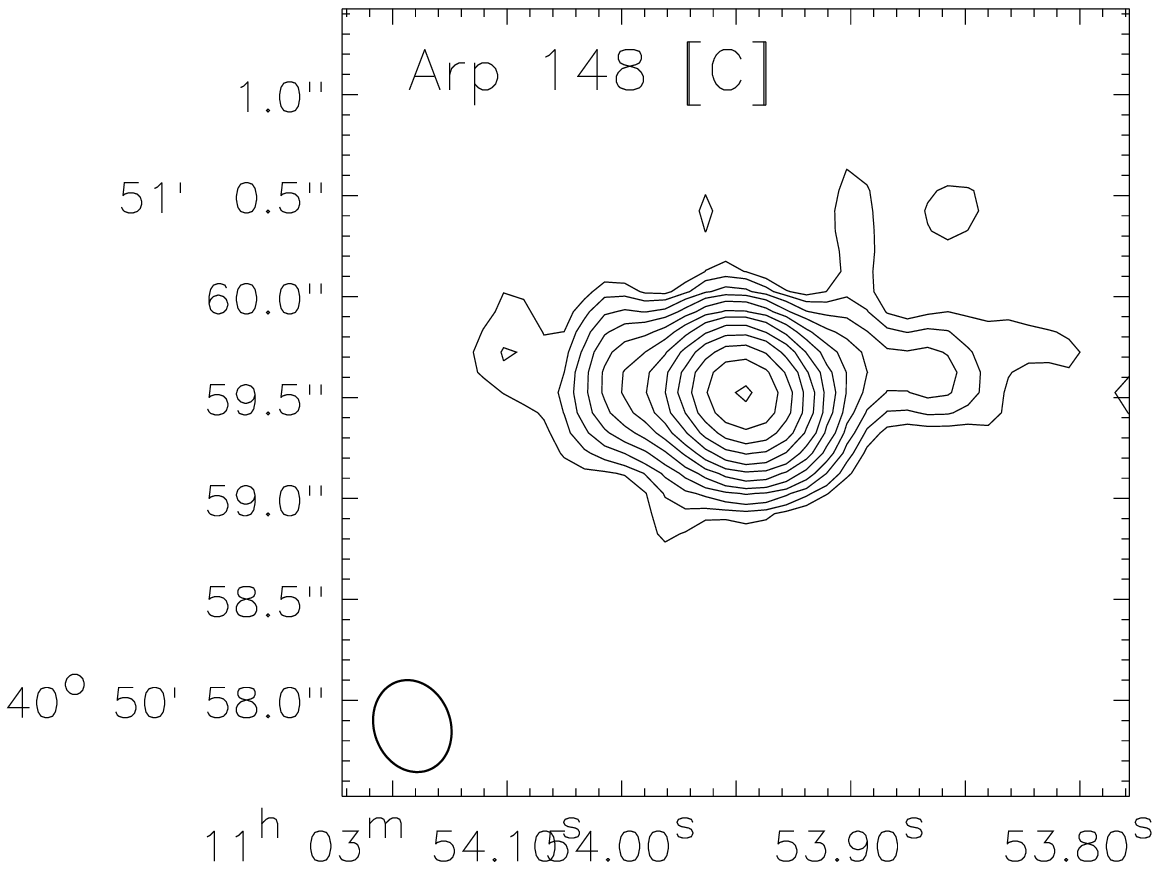}
\includegraphics[scale=0.38]{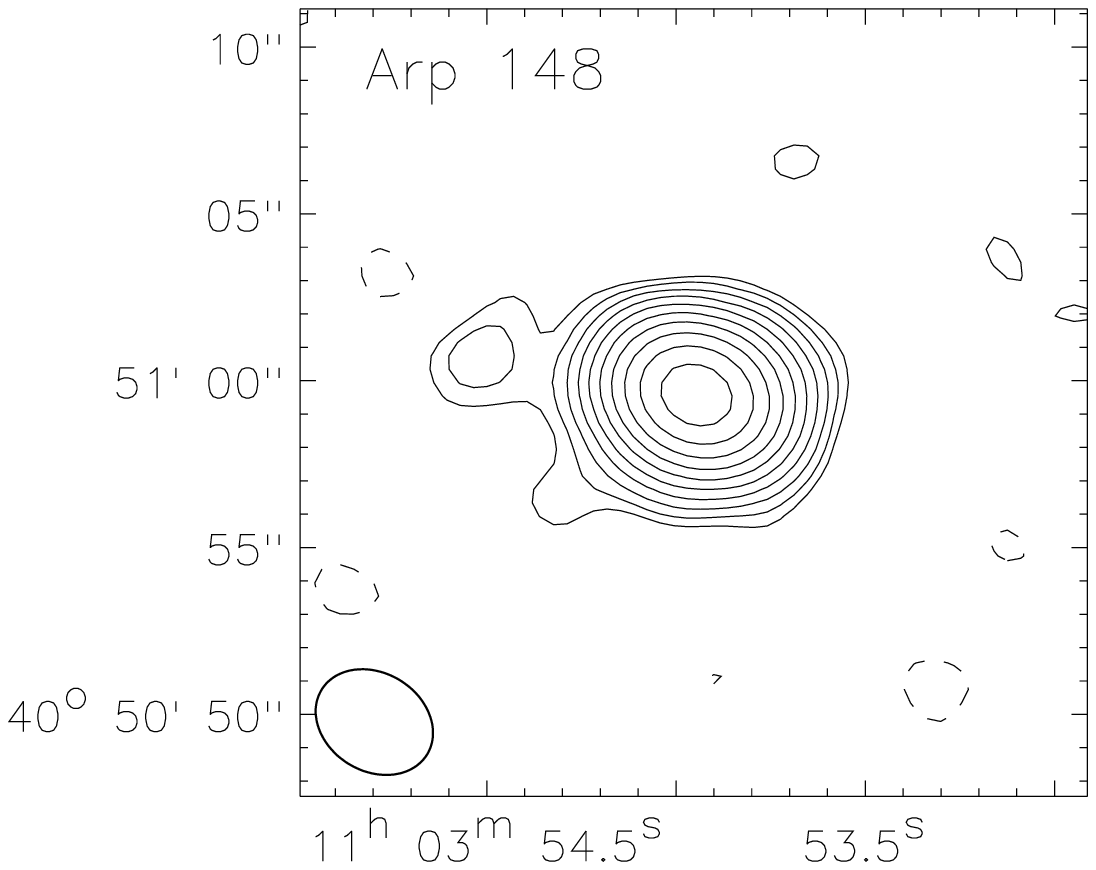}
}
\centerline{
\includegraphics[scale=0.38]{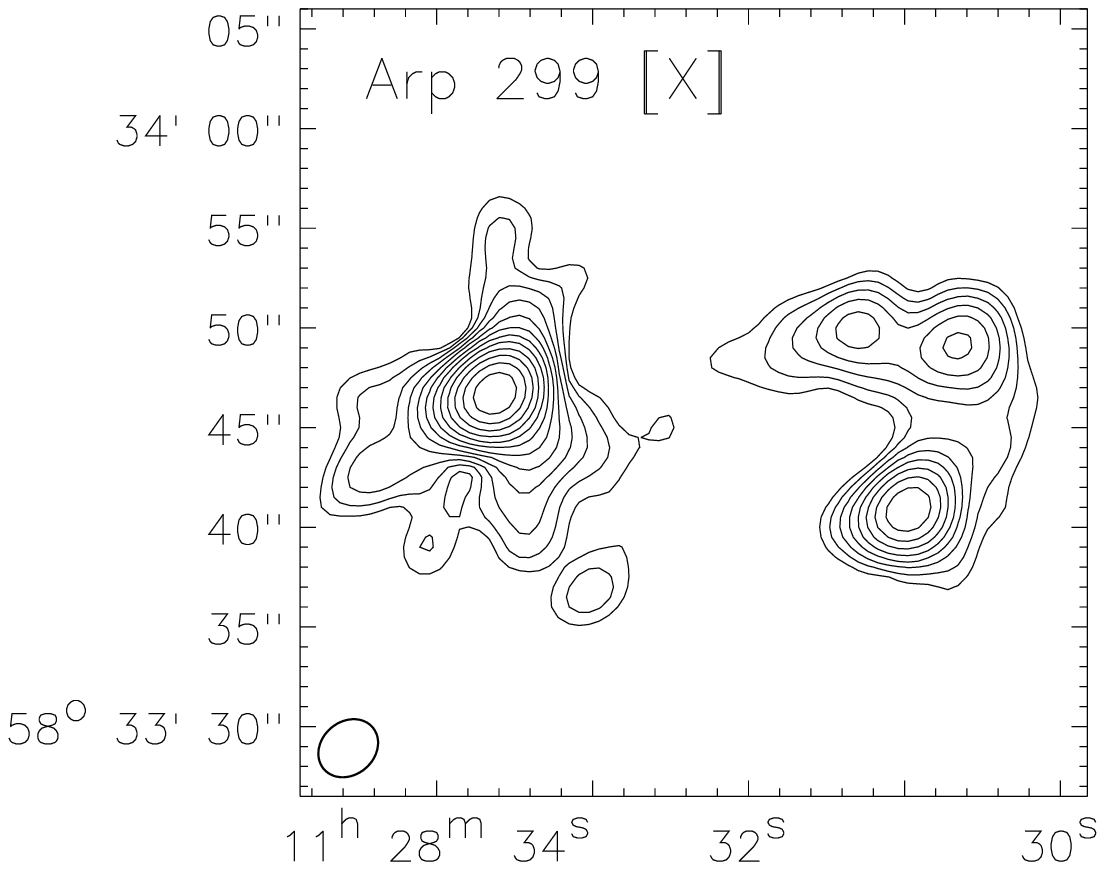}
\includegraphics[scale=0.38]{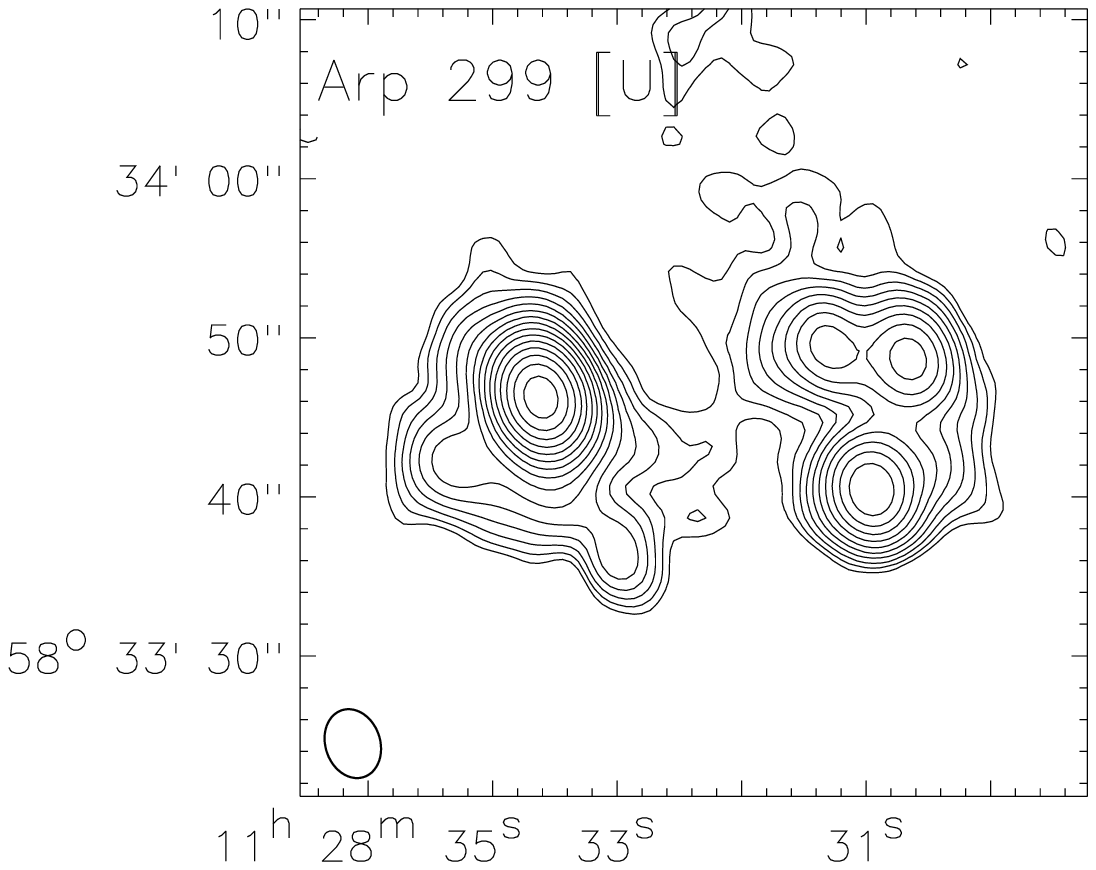}
\includegraphics[scale=0.38]{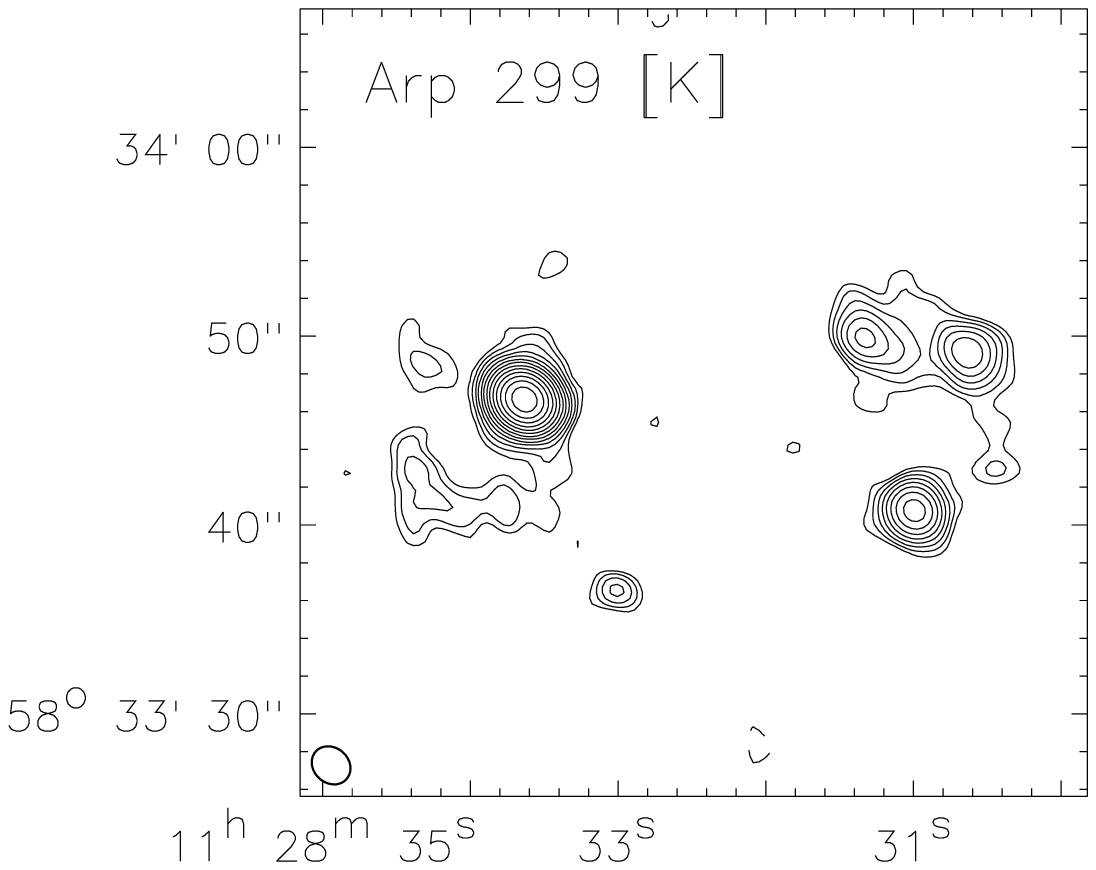}
\includegraphics[scale=0.38]{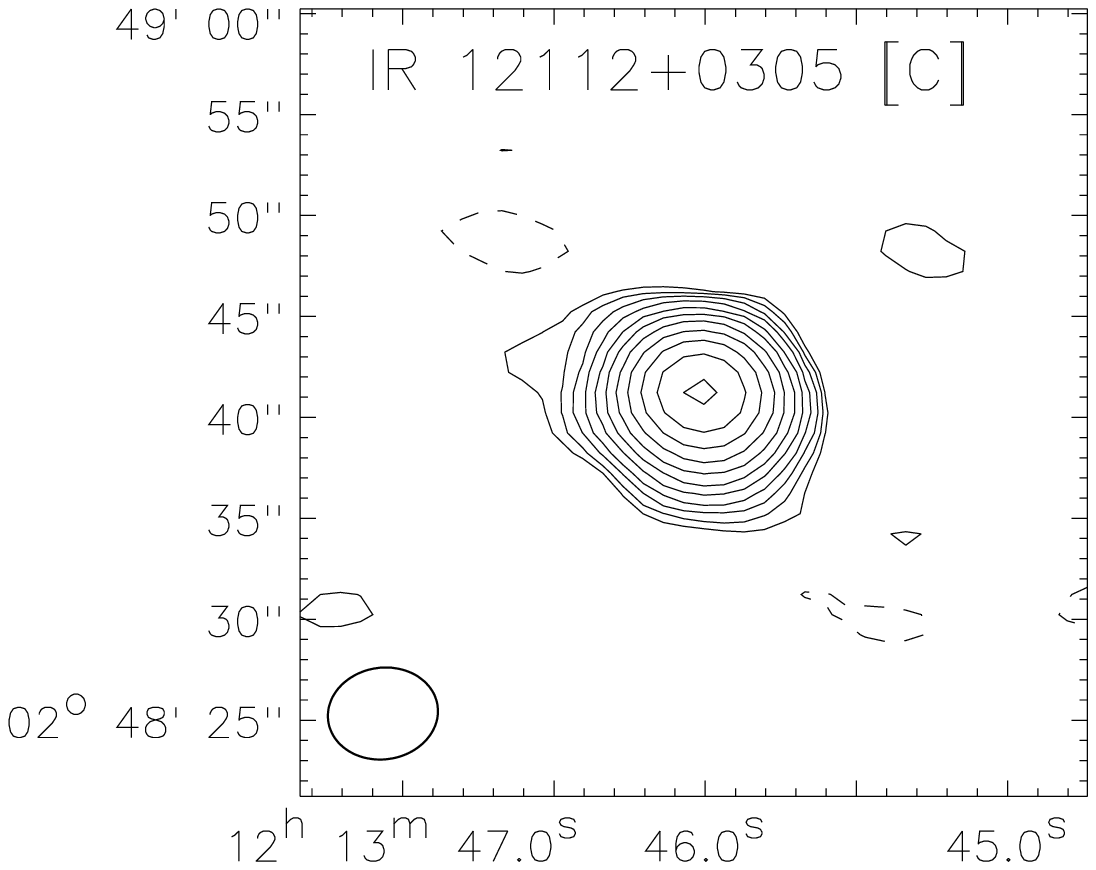}
}
\centerline{
\includegraphics[scale=0.38]{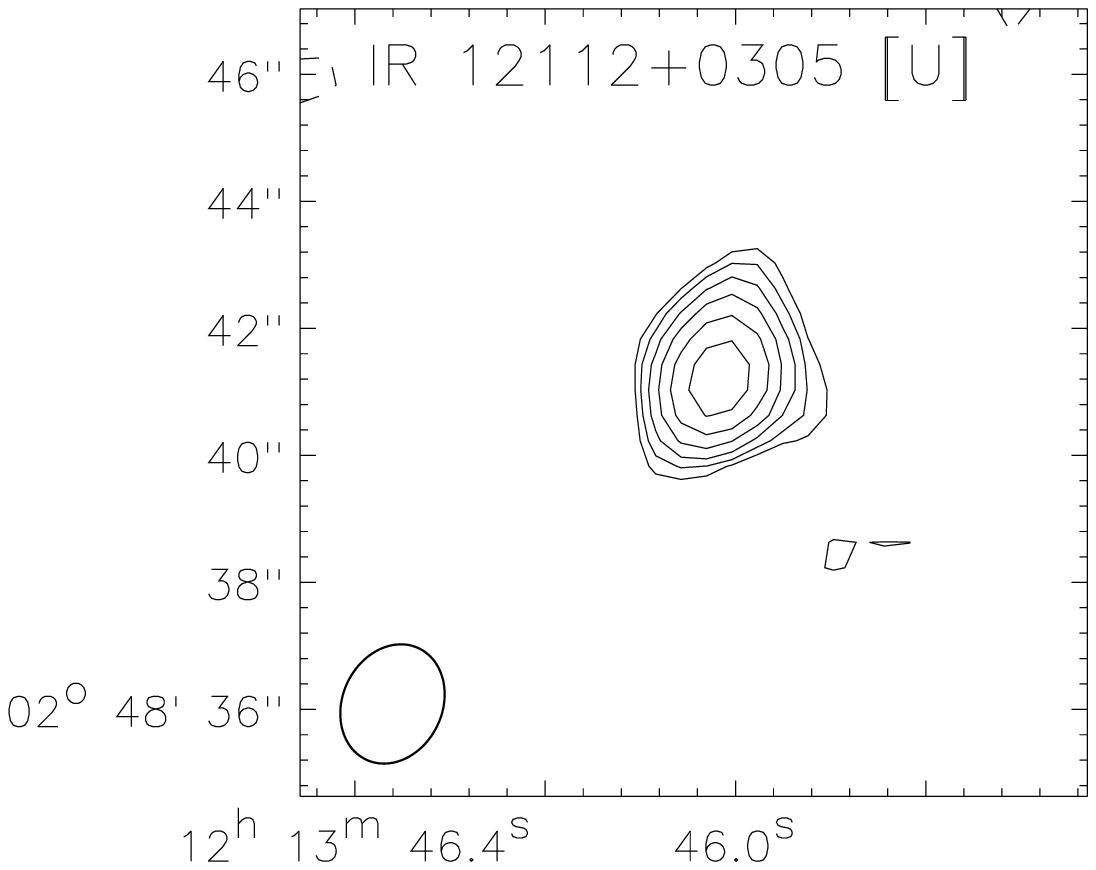}
\includegraphics[scale=0.38]{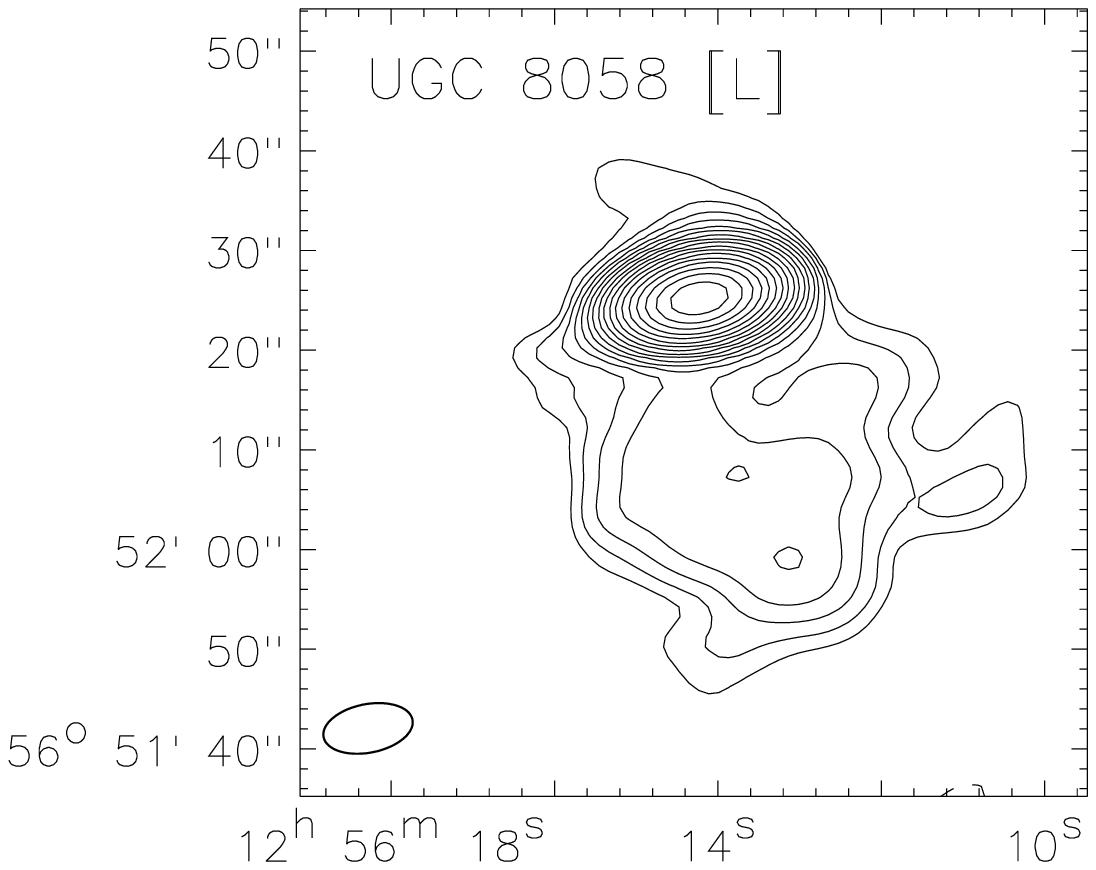}
\includegraphics[scale=0.38]{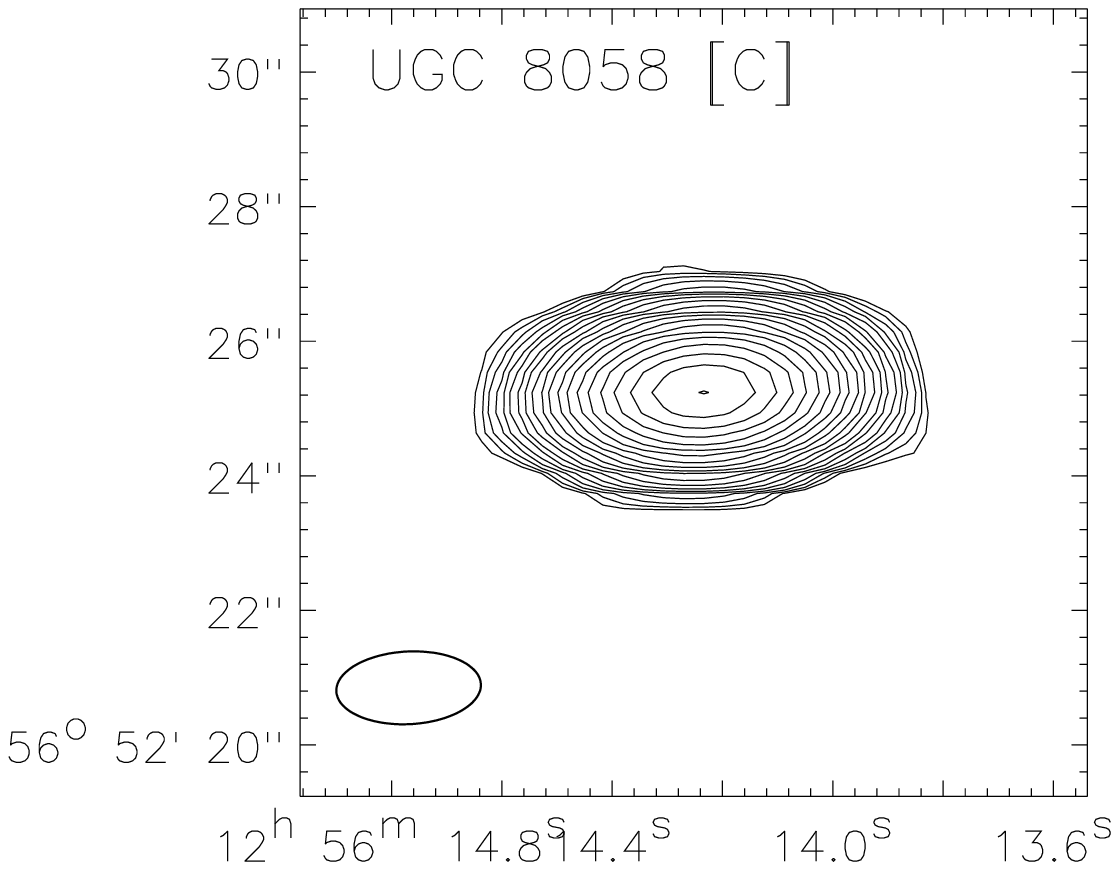}
\includegraphics[scale=0.38]{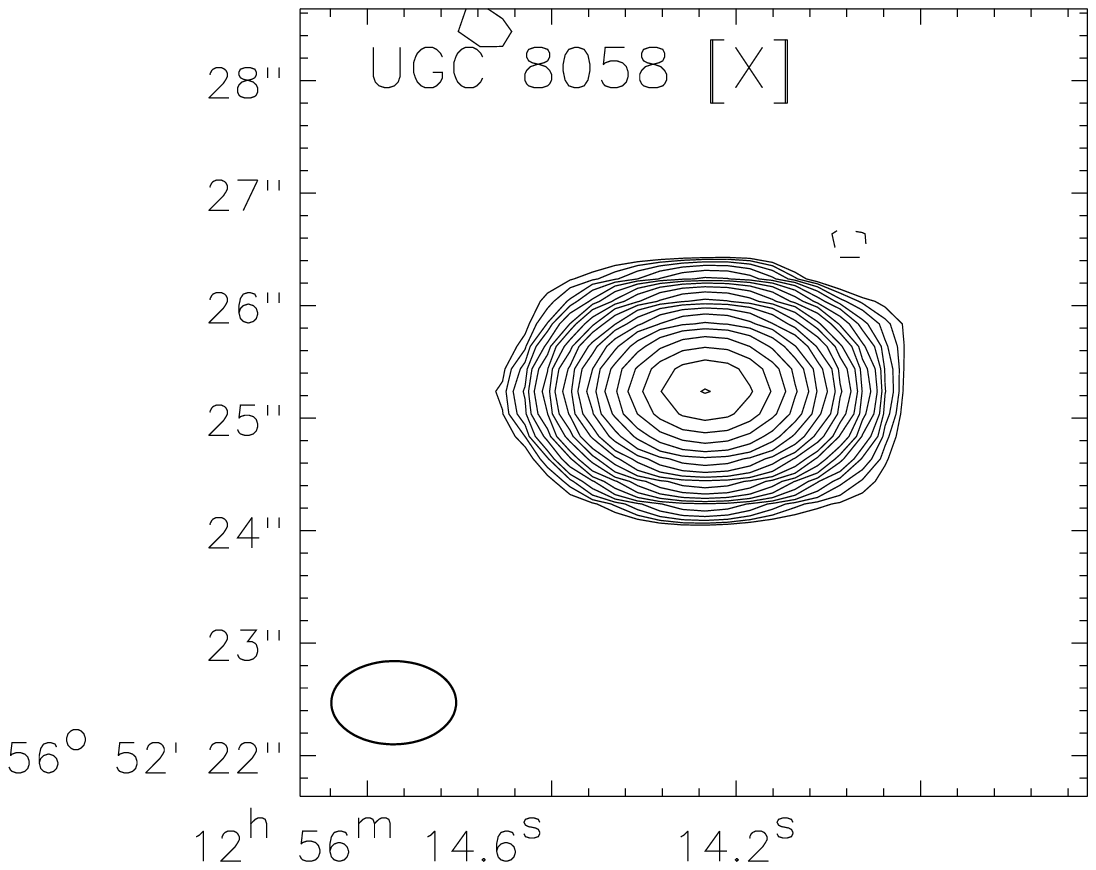}
}
\centerline{
\includegraphics[scale=0.38]{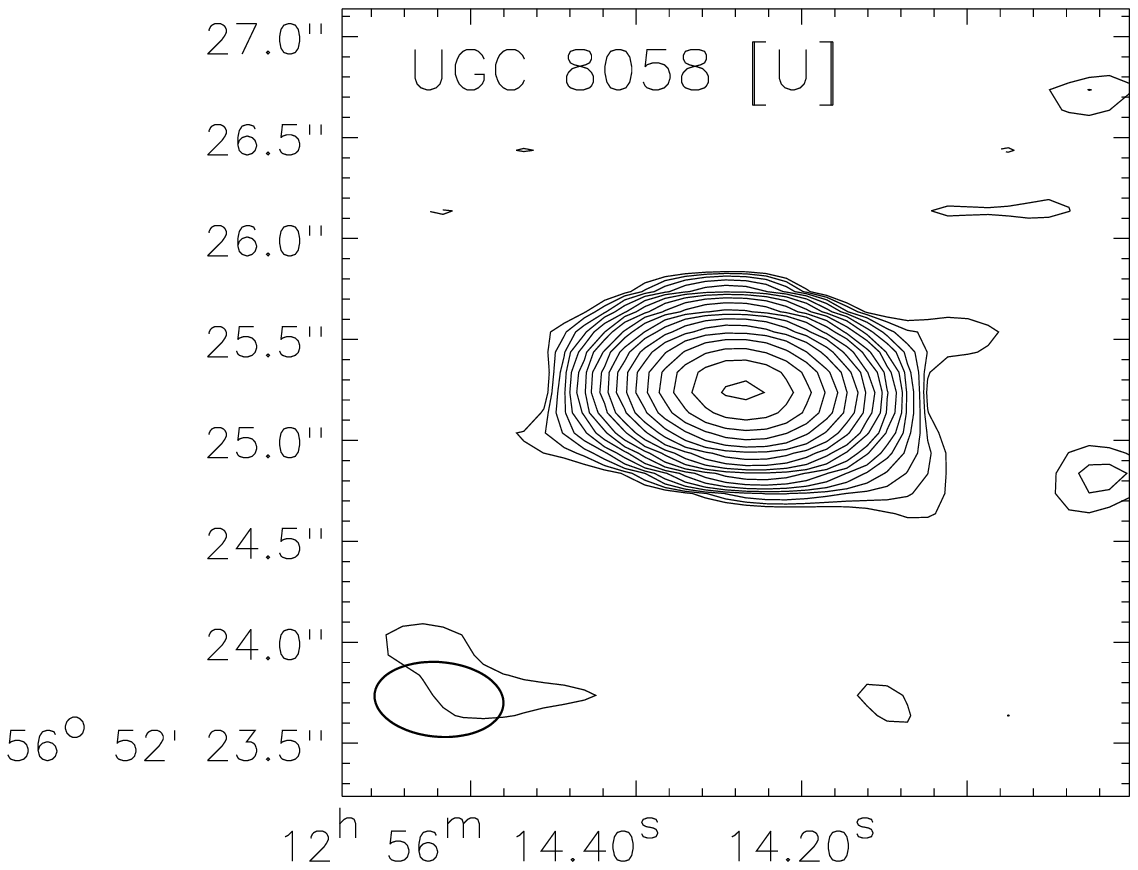}
\includegraphics[scale=0.38]{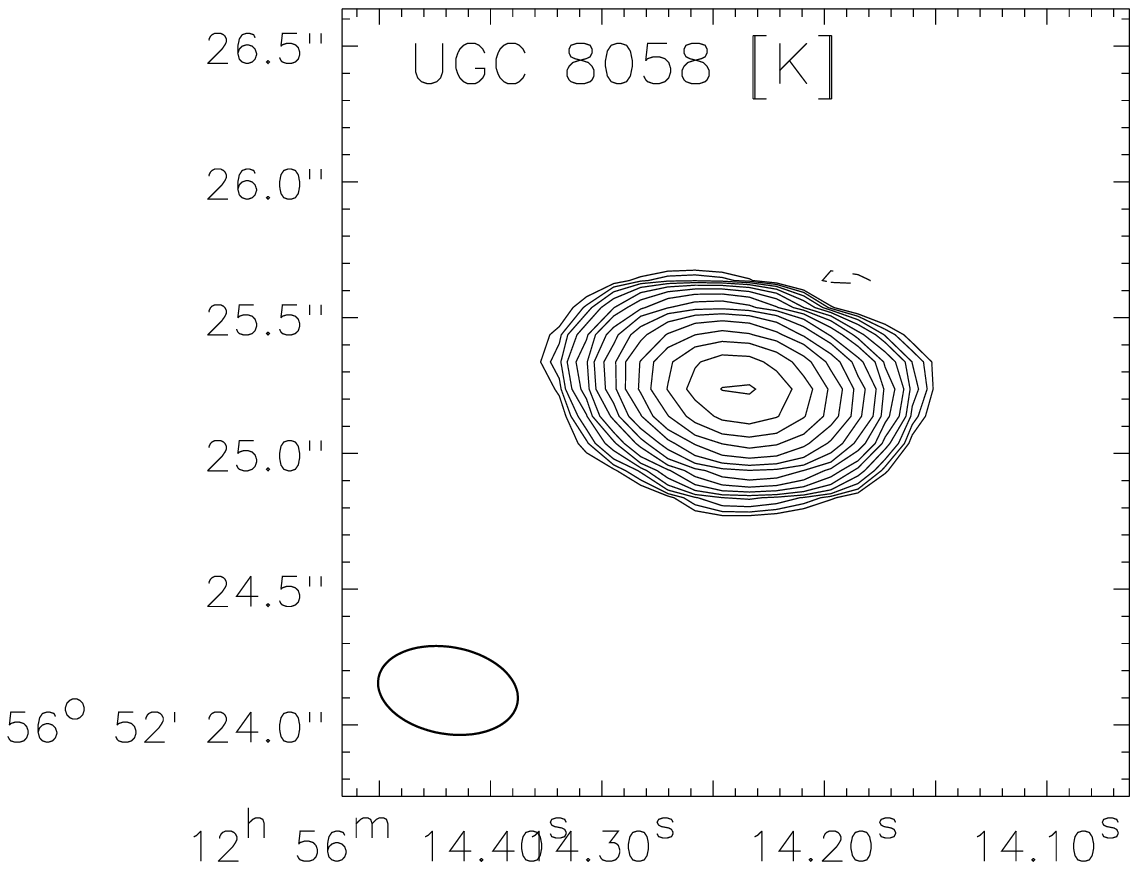}
\includegraphics[scale=0.38]{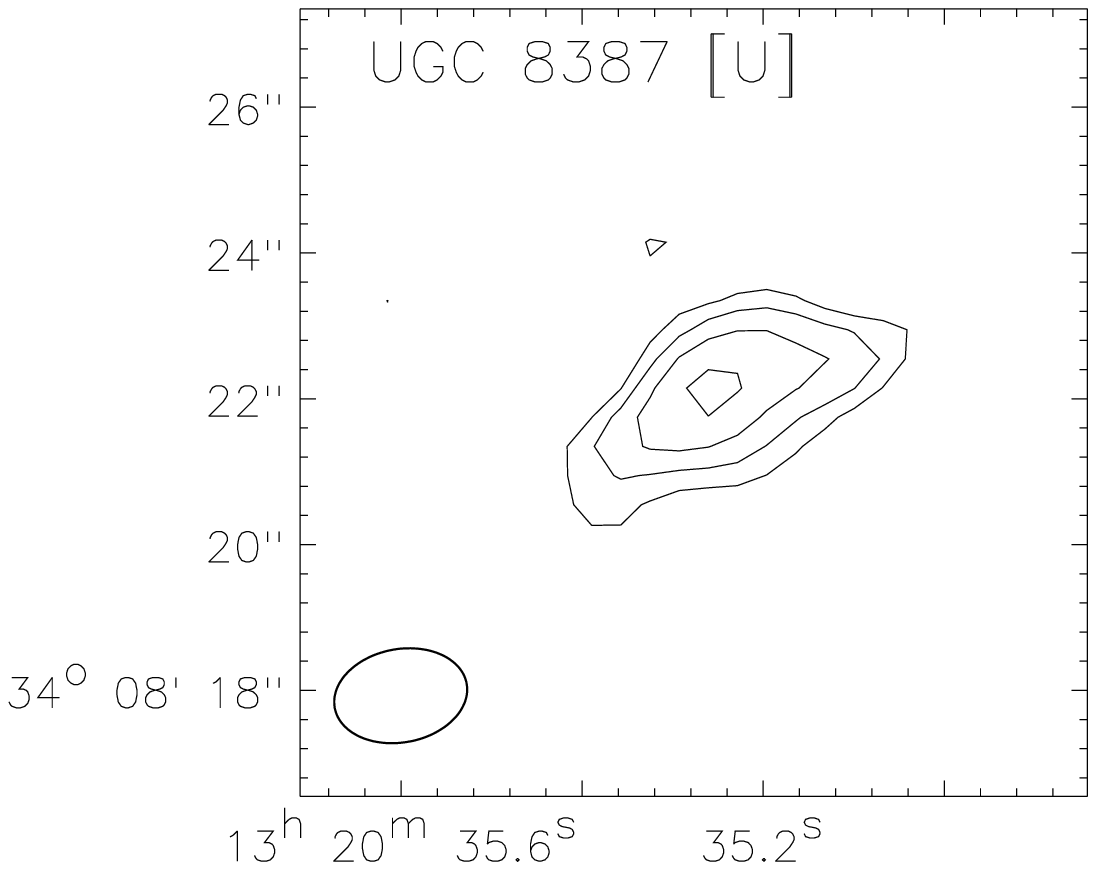}
\includegraphics[scale=0.38]{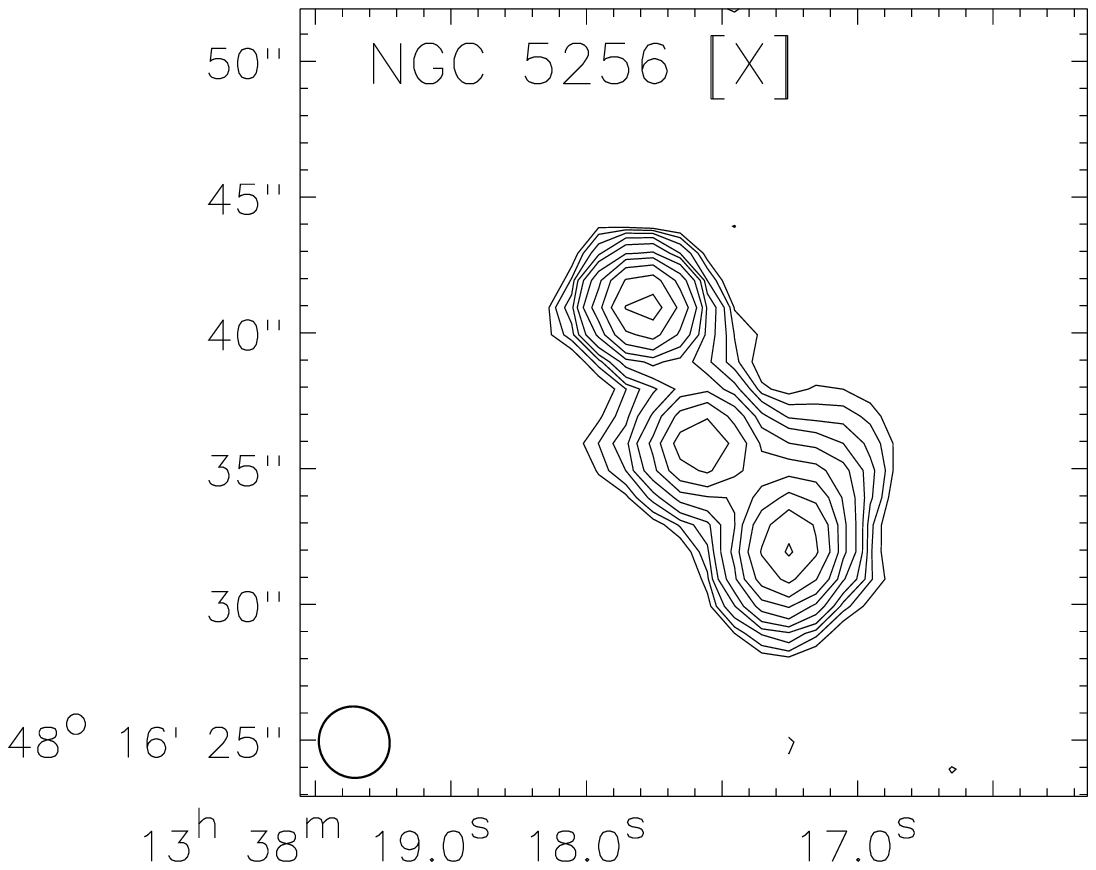}
}
\centerline{
\includegraphics[scale=0.38]{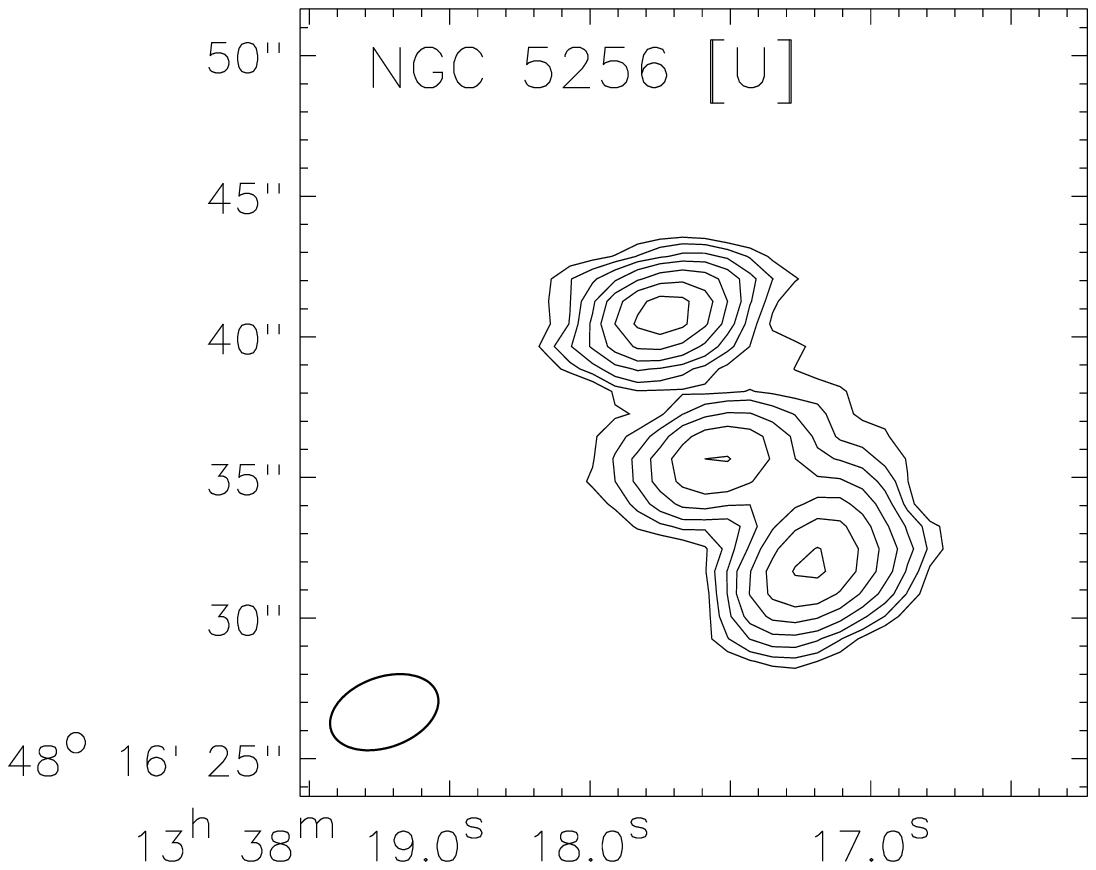}
\includegraphics[scale=0.38]{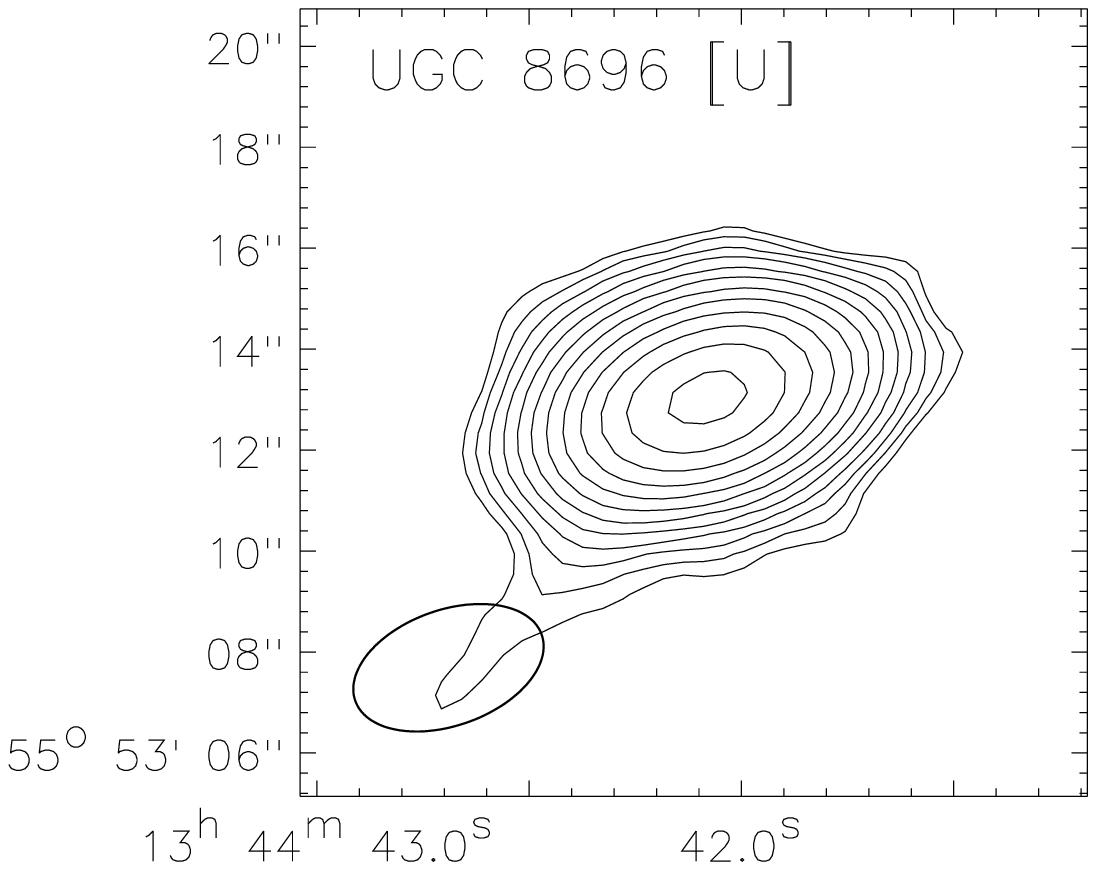}
\includegraphics[scale=0.38]{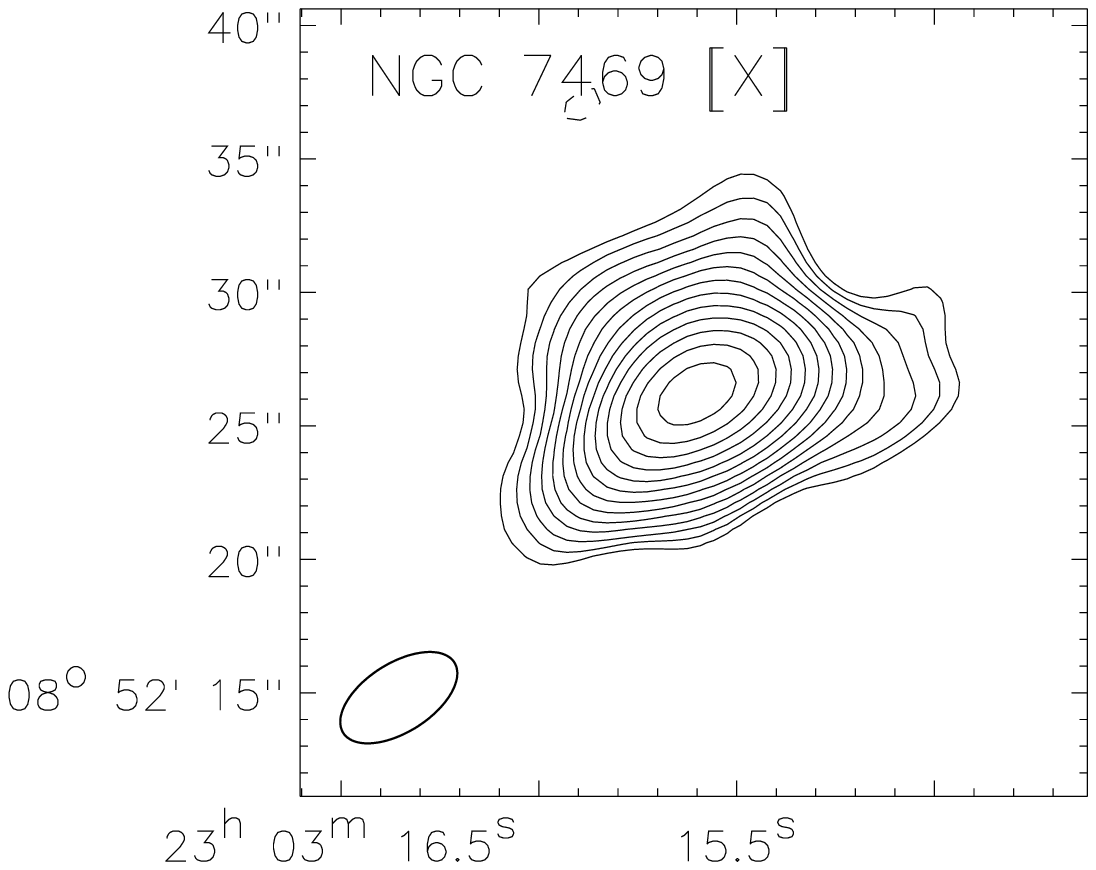}
}
\caption{Contour maps for the archival VLA data. In brackets after the name of the source is the name of the radio frequency band; L (1.4~GHz), C (4.8~GHz), X (8.4~GHz), U (15~GHz), K (22.5~GHz). Contour levels are $2^{-n\,x_{i}/2}$ where x = 0, 1, 2, \dots The dotted contour has the same absolute value as the first contour but is negative. The value of $n$ for each contour map is included in table~\ref{tab:archive}. The synthesized beam is shown in the  bottom left of each panel.}
\label{fig:archive_maps}
\end{figure*}

In addition to the above data and fluxes taken from the literature the
VLA archive was searched  for data at additional frequencies so as to
maximize the coverage of the radio spectrum. Data suitable  for the
measurement of integrated fluxes\footnote{i.e., where the source extent was 
smaller than the largest angular scale which can be mapped by the VLA 
in the given configuration and frequency band.} were downloaded and
reduced following standard procedures. Where the sources were of
sufficient strength, a single iteration of self-calibration was
applied allowing for the correction of antenna phases only. Fluxes
were measured by direct integration on the {\sc clean}ed maps and are 
presented in table~\ref{tab:archive}. For UGC~08058
(Mrk~231) archival data were downloaded and reduced, despite the
presence of measured fluxes in the literature, because of evidence
that the radio emission from this source is variable. A significant fraction 
of its radio flux must therefore come from an AGN. The 5 fluxes
listed in table ~\ref{tab:archive} were all measured on the same date.
The maps of sources reduced from the VLA archive are shown in 
Fig.~\ref{fig:archive_maps}.

All calibration and mapping was carried using the NRAO Astonomical Image Processing System ({\sc aips}). The radio fluxes for our sample galaxies (in the range 100~MHz to 100~GHz) are listed in table 1 in the appendix, along with notes on selected sources.

\subsection{Data at other frequencies}
In order to make comparisons between our radio data and infrared data,
we also collected J,H and K-band fluxes from the 2MASS Extended Source
Catalogue (Jarrett et al., 2000) and  12, 25, 60 and $100\;\rm \mu m$
fluxes from the IRAS Faint Source Catalogue (Moshir et al., 1990).

In the collection of fluxes at other frequencies from the literature,
care was taken to ensure that the fluxes were not confused or, in the case
of interferometric radio data, underestimated due to high resolution. 
While most sources are compact and do not have other strong sources nearby, 
a small number of sources required corrections to some fluxes.

Confusion is most likely to be a problem for the IRAS fluxes due to the very low
resolution. To try to quantify any possible problems with confusion in the absence of a
higher resolution FIR survey, we used the NVSS and the far infrared-radio correlation.
In those cases where an unrelated radio source was within the IRAS beam we reduced the
IRAS flux by the fractional contribution of this source in the radio. Few sources
were affected, but one source (IRAS~03359+1523) has its FIR emission confused with a source 
of similar strength. Because the error on a correction of this scale would render the 
fluxes unreliable this source will not be considered in the model fitting described in paper 2. 
The ULIRGs for which the IRAS fluxes were reduced because of the presence of a nearby 
contaminating source were as follows: UGC~6436 (26\%), IRAS~12112+0305 (15\%), NGC~6286 (6\%),
UGC~5101 (8\%).

The full list of fluxes for the sample galaxies from the near-infrared to the radio will be presented in paper 2, where they are used to model the spectral energy distributions.

\begin{table*}
\caption{22.5~GHz fluxes derived from the new VLA observations. $^*$ Observations described in Prouton et al. (2004)}
\centering
\begin{tabular}{c c c c c c}
\hline\hline
Source & R.A. & Dec. & Source flux & Phase cal & Phase cal flux \\
   & (J2000) & (J2000) & (mJy) & (J2000) & (Jy)  \\
\hline
NGC~34$^{*}$        & 00 11 06.550 & -12 06 26.33 & $7.41 \pm 0.14$ & 2358-103 & $0.5719 \pm 0.005$  \\
IC~1623             & 01 07 47.49  & -17 30 26.3  & $21.7 \pm 2.0$  & 0050-094 & $0.805 \pm 0.007$   \\
CGCG436-30$^{*}$    & 01 20 02.722 & +14 21 42.94 & $8.73 \pm 0.31$ & 0121+118 & $1.177 \pm 0.007$   \\
IRAS~01364-1042$^{*}$ & 01 38 52.870 & -10 27 11.70 & $3.97 \pm 0.19$ & 0141-094 & $0.7395 \pm 0.003$  \\
IIIZw~35            & 01 44 30.53  & +17 06 08.5  & $9.7 \pm 0.3$   & 0121+118 & $1.449 \pm 0.011$   \\
UGC~2369            & 02 54 01.81  & +14 58 14.5  & $5.3 \pm 0.3$   & 0238+166 & $2.041 \pm 0.015$   \\
IRAS~03359+1523       & 03 38 47.12  & +15 32 53.5  & $4.5 \pm 0.3$   & 0325+226 & $0.768 \pm 0.006$   \\
NGC~1614            & 04 34 00.02  & -08 34 44.98 & $21.0 \pm 1$    & 0423-013 & $7.480 \pm 0.058$   \\
NGC~2623            & 08 38 24.08  & +25 45 16.40 & $18.3 \pm 0.2$  & 0830+241 & $1.822 \pm 0.013$   \\
IRAS~08572+3915$^{*}$ & 09 00 25.390 & +39 03 54.40 & $3.18 \pm 0.3$  & 0927+390 & $7.688 \pm 0.15$    \\
UGC~4881            & 09 15 55.52  & +44 19 57.5  & $3.5 \pm 0.5$   & 0920+446 & $2.140 \pm 0.010$   \\
UGC~5101            & 09 35 51.72  & +61 21 11.29 & $18.0 \pm 1.0$  & 0958+655 & $0.972 \pm 0.010$   \\
IRAS~10173+0828       & 10 20 00.22  & +08 13 34.5  & $2.9 \pm 0.4$   & 1041+061 & $1.672 \pm 0.009$   \\
Arp~148             & 11 03 53.95  & +40 50 59.5  & $6.1  \pm 0.3$  & 1146+399 & $1.330 \pm 0.012$   \\
UGC~6436            & 11 25 45.15  & +14 40 29.4  & $2.8  \pm 0.5$  & 1118+125 & $1.238 \pm 0.016$   \\
IRAS~12112+0305       & 12 13 46.1   & +02 48 41.5  & $5.7 \pm 0.3$   & 1224+035 & $0.788 \pm 0.006$   \\
UGC~8387            & 13 20 35.34  & +34 08 22.19 & $7.5 \pm 2.0$   & 1310+323 & $0.963 \pm 0.042$   \\
UGC~8696            & 13 44 42.11  & +55 53 13.15 & $14.3 \pm 0.3$  & 1419+543 & $0.935 \pm 0.023$   \\
IRAS~14348-1447       & 14 37 38.30  & -15 00 24.0  & $4.5 \pm 0.5$   & 1448-163 & $0.376 \pm 0.006$   \\
IZw~107             & 15 18 06.07  & +42 44 45.19 & $3.6  \pm 0.3$  & 1521+436 & $0.195 \pm 0.005$   \\
IRAS~15250+3609       & 15 26 59.40  & +35 58 37.53 & $4.2 \pm 0.3$   & 1521+436 & $0.195 \pm 0.005$   \\
NGC~6286            & 16 58 31.70  & +58 56 14.50 & $12.2 \pm 1.5$  & 1638+573 & $1.259 \pm 0.018$   \\
NGC~7469$^{*}$      & 23 03 15.623 & +08 52 26.39 & $17.50 \pm 0.50$& 2320+052 & $0.5661 \pm 0.004$  \\
IC~5298$^{*}$       & 23 16 00.690 & +25 33 24.20 & $3.86 \pm 0.18$ & 2321+275 & $0.7178 \pm 0.009$  \\
Mrk~331$^{*}$       & 23 51 26.802 & +20 35 09.87 & $9.95 \pm 0.29$ & 2358+199 & $0.2683 \pm 0.002$  \\

\hline
\end{tabular}
\label{tab:cal}
\end{table*}

\begin{table}
\caption{8.4~GHz fluxes for the newly observed objects.}
\centering
\begin{tabular}{c c c c}
\hline\hline
Source & Source flux & Phase cal & Phase cal flux \\
   & (mJy) & (J2000) & (Jy) \\
\hline
IC~1623       & $ 54.6 \pm 3.0 $ & 0118-216 & $ 0.854 \pm 0.002 $ \\
Arp~148       & $ 11.8 \pm 0.6 $ & 1130+382 & $ 1.338 \pm 0.005 $ \\
UGC~6436      & $  5.2 \pm 0.3 $ & 1122+180 & $ 0.639 \pm 0.003 $ \\
IRAS~14348-1447 & $ 10.8 \pm 0.5 $ & 1439-155 & $ 0.509 \pm 0.003 $ \\
IZw~107       & $ 14.3 \pm 0.7 $ & 1500+478 & $ 0.341 \pm 0.001 $ \\
NGC~6286      & $ 31.5 \pm 1.6 $ & 1638+573 & $ 1.261 \pm 0.003 $ \\
Mrk~331       & $ 22.3 \pm 1.0 $ & 0010+174 & $ 0.670 \pm 0.002 $ \\
\hline
\end{tabular}
\label{tab:8GHz}
\end{table}

\begin{table*}
\caption{Fluxes derived from archival VLA data. $^*$ No flux calibrator observed. Flux of 0513-219 set to this value using the VLA calibrator database. The last column, $n$, shows the scaling of the contours in Fig.~\ref{fig:archive_maps}.}
\centering
\begin{tabular}{lcccccccc}
\hline\hline
Source & Frequency & Flux & Phase cal & Phase cal flux & Program ID & Array & Date & $n$\\
   & (GHz) & (mJy) & (J2000) & (Jy) & & & &\\
\hline
         NGC~34 & 4.8  & $28.4 \pm 0.3$	& 2358-103 & $0.8080 \pm 0.0013$  & AM451  & B  & 1994-Jul-24 & 6 \\
       IIIZw~35 & 15   & $10.0 \pm 1$   & 0149+189 & $0.3538 \pm 0.003$   & TYP04  & B  & 2002-sep-06 & -1\\
       NGC~1614 & 8.4  & $40.7 \pm 0.3$ & 0423-013 & $3.1178 \pm 0.004$   & AK331  & C  & 1993-aug-03 & 6 \\
IRAS~05189-2524 & 15   & $7.8 \pm 0.2$  & 0513-219 & $0.92   \pm 0.01^{*}$& AC624  & C  & 2003-jan-05 & 4 \\
IRAS~10173+0828 & 4.8  & $5.88 \pm 0.2$ & 1150-003 & $1.7056 \pm 0.004$   & AK184  & CD & 1988-mar-28 & 4 \\
	        & 15   & $3.1 \pm 0.3$  & 1150-003 & $0.9135 \pm 0.004$   & AK184  & CD & 1988-mar-28 & 1 \\ 
        Arp~148 & 4.8  & $14.0 \pm 0.5$ & 1130+382 & $0.7813 \pm 0.002$   & AH333  & A  & 1989-jan-09 & 5 \\
	        & 15   & $8.0 \pm 1$    & 1130+382 & $0.6237 \pm 0.007$   & AJ105  & C  & 1984-may-24 & 4 \\
        Arp~299 & 8.4  & $243 \pm 8$    & 1219+484 & $0.7873 \pm 0.003$   & AW641  & C  & 2005-aug-20 & -1\\
	        & 15   & $160 \pm 3$    & 1219+484 & $0.8489 \pm 0.012$   & AS568  & C  & 1999-jan-21 & 2 \\
        	& 22.5 & $74 \pm 2$     & 1128+594 & $0.4509 \pm 0.004$   & AN095  & C  & 2001-aug-25 & 2 \\
IRAS~12112+0305 & 4.8  & $14.7 \pm 0.2$ & 1150-003 & $1.7056 \pm 0.004$   & AK184  & CD & 1988-mar-28 & 3 \\
	        & 15   & $5.7 \pm 0.2$  & 1150-003 & $0.9135 \pm 0.004$   & AK184  & CD & 1988-mar-28 & 1 \\
      UGC~08058 & 1.4  & $274 \pm 3$    & 1219+484 & $0.5987 \pm 0.001$   & BU006Y & B  & 1995-nov-17 & 1 \\
	        & 4.8  & $265 \pm 1$    & 1219+484 & $0.7885 \pm 0.002$   & BU006Y & B  & 1995-nov-17 & 3 \\
	        & 8.4  & $189 \pm 1$    & 1219+484 & $0.9363 \pm 0.002$   & BU006Y & B  & 1995-nov-17 & 4 \\
		& 15   & $146 \pm 1$    & 1219+484 & $1.0916 \pm 0.005$   & BU006Y & B  & 1995-nov-17 & 3 \\
		& 22.5 & $137 \pm 2$    & 1219+484 & $1.1123 \pm 0.008$   & BU006Y & B  & 1995-nov-17 & 0 \\
       UGC~8387 & 15   & $12 \pm 2$     & 3C286	   & $3.45375$		  & AN35   & C  & 1985-aug-30 & 0 \\
       NGC~5256 & 8.4  & $24.1 \pm 1$   & 1417+461 & $0.5880 \pm 0.003$	  & AP233  & C  & 1992-apr-28 & 4 \\
                & 15   & $21.0 \pm 1$   & 1349+536 & $0.6003 \pm 0.006$	  & AM198  & CD & 1987-jan-22 & 2 \\
       UGC~8696 & 15   & $29.5 \pm 1$   & 1349+536 & $0.6003 \pm 0.006$	  & AM198  & CD & 1987-jan-22 & 2 \\
       NGC~7469 & 8.4  & $50.2 \pm 0.6$ & 2320+052 & $0.5897 \pm 0.05$	  & AP233  & C  & 1992-apr-28 & 3 \\
\hline
\end{tabular}
\label{tab:archive}
\end{table*}

\begin{landscape}
\begin{table*}
\caption{Far-infrared-radio flux ratios, $q$, at 1.4 and 8.4~GHz and radio spectral indices for the whole sample in various frequency ranges. The value of ${\rm FIR} = 1.26\times 10^{-14}(2.58\,S_{60} + S_{100})$, as used to calculate $q$, is also given.}
\hspace{-8cm}
\begin{tabular}{lcccccccccc}
\hline\hline
Name & FIR & $q_{1.4}$ & $q_{8.4}$ & $\alpha^{1.4}_{22.5}$ & $\alpha^{1.4}_{4.8}$ & $\alpha^{1.4}_{8.4}$ & $\alpha^{4.8}_{8.4}$ & $\alpha^{8.4}_{15}$ & $\alpha^{8.4}_{22.5}$ & $\alpha^{15}_{22.5}$\\
& $(/10^{-13}\,\rm W\,m^{-2})$ & & & & & & & & &\\
\hline
   NGC~34    & $7.69\pm 0.02$ & $2.49\pm 0.02$ & $3.13\pm 0.02$ & $-0.79\pm 0.02$ & $-0.70\pm 0.03$ & $-0.83\pm 0.04$ & $-1.12\pm 0.10$ & $      -      $ & $-0.73\pm 0.06$ & $      -      $\\
  IC~1623    & $11.2\pm 0.63$ & $2.08\pm 0.03$ & $2.74\pm 0.03$ & $-0.88\pm 0.04$ & $-0.77\pm 0.11$ & $-0.85\pm 0.02$ & $-1.01\pm 0.23$ & $      -      $ & $-0.94\pm 0.10$ & $      -      $\\
CGCG~436-30  & $4.70\pm 0.03$ & $2.40\pm 0.01$ & $2.99\pm 0.02$ & $-0.63\pm 0.02$ & $-0.69\pm 0.05$ & $-0.77\pm 0.03$ & $-0.94\pm 0.12$ & $      -      $ & $-0.38\pm 0.06$ & $      -      $\\
IRAS~0136-1042  & $3.02\pm 0.02$ & $2.71\pm 0.02$ & $2.99\pm 0.02$ & $-0.50\pm 0.02$ & $-0.24\pm 0.05$ & $-0.37\pm 0.04$ & $-0.65\pm 0.12$ & $      -      $ & $-0.74\pm 0.07$ & $      -      $\\
 IIIZw~35    & $5.74\pm 0.35$ & $2.58\pm 0.03$ & $2.89\pm 0.03$ & $-0.52\pm 0.02$ & $-0.35\pm 0.05$ & $-0.40\pm 0.03$ & $-0.52\pm 0.13$ & $-1.17\pm 0.19$ & $-0.72\pm 0.06$ & $-0.08\pm 0.26$\\
 UGC~2369    & $3.94\pm 0.16$ & $2.33\pm 0.02$ & $2.90\pm 0.03$ & $-0.80\pm 0.02$ & $      -      $ & $-0.73\pm 0.03$ & $      -      $ & $      -      $ & $-0.93\pm 0.08$ & $      -      $\\
IRAS~03359+1523 & $2.88\pm 0.14$ & $2.60\pm 0.03$ & $2.84\pm 0.03$ & $-0.53\pm 0.03$ & $-0.41\pm 0.05$ & $-0.32\pm 0.03$ & $-0.11\pm 0.13$ & $      -      $ & $-0.91\pm 0.08$ & $      -      $\\
  NGC~1614   & $14.6\pm 0.45$ & $2.45\pm 0.02$ & $2.98\pm 0.01$ & $-0.68\pm 0.02$ & $-0.63\pm 0.14$ & $-0.68\pm 0.02$ & $-0.78\pm 0.31$ & $      -      $ & $-0.67\pm 0.05$ & $      -      $\\
IRAS~05189-2524 & $5.89\pm 0.19$ & $2.74\pm 0.02$ & $3.14\pm 0.03$ & $    -        $ & $-0.43\pm 0.05$ & $-0.52\pm 0.03$ & $-0.71\pm 0.13$ & $-0.65\pm 0.10$ & $      -      $ & $      -      $\\
 NGC~2623    & $11.0\pm 0.35$ & $2.49\pm 0.02$ & $2.92\pm 0.03$ & $-0.60\pm 0.01$ & $-0.39\pm 0.14$ & $-0.55\pm 0.03$ & $-0.91\pm 0.32$ & $      -      $ & $-0.67\pm 0.05$ & $      -      $\\
IRAS~08572+3915 & $2.97\pm 0.02$ & $3.27\pm 0.04$ & $3.29\pm 0.02$ & $-0.11\pm 0.05$ & $ 0.02\pm 0.08$ & $-0.03\pm 0.06$ & $-0.13\pm 0.12$ & $      -      $ & $-0.26\pm 0.11$ & $      -      $\\
   UGC~4881  & $3.22\pm 0.13$ & $2.37\pm 0.02$ & $2.99\pm 0.03$ & $-0.85\pm 0.05$ & $      -      $ & $-0.80\pm 0.03$ & $      -      $ & $      -      $ & $-0.94\pm 0.15$ & $      -      $\\
   UGC~5101  & $6.28\pm 0.29$ & $1.99\pm 0.03$ & $2.50\pm 0.03$ & $-0.81\pm 0.02$ & $-0.69\pm 0.17$ & $-0.66\pm 0.03$ & $-0.59\pm 0.38$ & $-0.97\pm 0.58$ & $-1.09\pm 0.08$ & $-1.26\pm 0.83$\\
IRAS~10173+0828 & $2.58\pm 0.14$ & $2.84\pm 0.05$ & $3.10\pm 0.03$ & $-0.44\pm 0.06$ & $-0.42\pm 0.08$ & $-0.34\pm 0.06$ & $-0.15\pm 0.11$ & $-0.96\pm 0.19$ & $-0.63\pm 0.15$ & $-0.16\pm 0.42$\\
IRAS~10565+2448 & $5.84\pm 0.22$ & $2.44\pm 0.02$ & $3.04\pm 0.03$ & $    -        $ & $-0.77\pm 0.03$ & $-0.78\pm 0.03$ & $-0.81\pm 0.09$ & $      -      $ & $      -      $ & $      -      $\\
  Arp~148    & $3.23\pm 0.13$ & $2.37\pm 0.02$ & $2.87\pm 0.04$ & $-0.64\pm 0.02$ & $-0.78\pm 0.04$ & $-0.63\pm 0.05$ & $-0.31\pm 0.16$ & $-0.66\pm 0.26$ & $-0.67\pm 0.10$ & $-0.67\pm 0.33$\\
   UGC~6436  & $3.23\pm 0.06$ & $2.65\pm 0.02$ & $3.22\pm 0.03$ & $-0.70\pm 0.07$ & $      -      $ & $-0.73\pm 0.04$ & $      -      $ & $      -      $ & $-0.63\pm 0.19$ & $      -      $\\
  Arp~299    & $47.2\pm 2.13$ & $2.26\pm 0.03$ & $2.71\pm 0.02$ & $-0.80\pm 0.02$ & $-0.43\pm 0.10$ & $-0.58\pm 0.03$ & $-0.90\pm 0.21$ & $-0.72\pm 0.07$ & $-1.21\pm 0.04$ & $-1.90\pm 0.08$\\
IRAS~12112+0305 & $4.02\pm 0.17$ & $2.66\pm 0.02$ & $3.03\pm 0.03$ & $-0.51\pm 0.02$ & $-0.37\pm 0.03$ & $-0.47\pm 0.03$ & $-0.69\pm 0.09$ & $-0.97\pm 0.11$ & $-0.57\pm 0.07$ & $-0.00\pm 0.16$\\
   UGC~8058  & $14.2\pm 0.54$ & $2.14\pm 0.02$ & $2.30\pm 0.02$ & $-0.25\pm 0.01$ & $-0.03\pm 0.01$ & $-0.21\pm 0.01$ & $-0.60\pm 0.01$ & $-0.45\pm 0.01$ & $-0.33\pm 0.02$ & $-0.16\pm 0.04$\\
   UGC~8387  & $8.18\pm 0.49$ & $2.32\pm 0.03$ & $2.80\pm 0.03$ & $-0.95\pm 0.10$ & $-0.78\pm 0.05$ & $-0.61\pm 0.03$ & $-0.24\pm 0.13$ & $-1.84\pm 0.30$ & $-1.56\pm 0.28$ & $-1.16\pm 0.78$\\
   NGC~5256  & $3.78\pm 0.17$ & $1.90\pm 0.03$ & $2.62\pm 0.03$ & $-0.77\pm 0.04$ & $-0.77\pm 0.06$ & $-0.93\pm 0.03$ & $-1.28\pm 0.14$ & $-0.23\pm 0.11$ & $-0.48\pm 0.11$ & $-0.85\pm 0.27$\\
   UGC~8696  & $9.75\pm 0.30$ & $2.25\pm 0.02$ & $2.78\pm 0.03$ & $-0.83\pm 0.01$ & $-0.58\pm 0.17$ & $-0.67\pm 0.03$ & $-0.88\pm 0.39$ & $-0.67\pm 0.10$ & $-1.13\pm 0.06$ & $-1.79\pm 0.10$\\
IRAS~14348-1447 & $3.12\pm 0.14$ & $2.37\pm 0.02$ & $2.89\pm 0.04$ & $-0.75\pm 0.04$ & $      -      $ & $-0.67\pm 0.05$ & $      -      $ & $      -      $ & $-0.89\pm 0.15$ & $      -      $\\
  IZw~107    & $4.34\pm 0.13$ & $2.36\pm 0.02$ & $2.91\pm 0.03$ & $-0.95\pm 0.03$ & $-0.37\pm 0.03$ & $-0.70\pm 0.04$ & $-1.44\pm 0.13$ & $      -      $ & $-1.40\pm 0.11$ & $      -      $\\
IRAS~15250+3609 & $3.11\pm 0.12$ & $2.76\pm 0.02$ & $2.90\pm 0.03$ & $-0.45\pm 0.03$ & $-0.09\pm 0.19$ & $-0.18\pm 0.04$ & $-0.38\pm 0.42$ & $-0.47\pm 0.65$ & $-0.93\pm 0.09$ & $-1.59\pm 0.94$\\
  Arp~220    & $47.9\pm 1.41$ & $2.59\pm 0.02$ & $2.94\pm 0.03$ & $-0.46\pm 0.03$ & $-0.36\pm 0.08$ & $-0.44\pm 0.03$ & $-0.63\pm 0.19$ & $-0.51\pm 0.09$ & $-0.50\pm 0.08$ & $-0.49\pm 0.17$\\
   NGC~6286  & $5.08\pm 0.13$ & $1.94\pm 0.02$ & $2.63\pm 0.02$ & $-0.95\pm 0.05$ & $-0.88\pm 0.13$ & $-0.90\pm 0.03$ & $-0.93\pm 0.28$ & $      -      $ & $-1.05\pm 0.15$ & $      -      $\\
   NGC~7469  & $13.3\pm 0.08$ & $2.29\pm 0.01$ & $2.85\pm 0.01$ & $-0.84\pm 0.01$ & $-0.77\pm 0.10$ & $-0.71\pm 0.02$ & $-0.59\pm 0.21$ & $      -      $ & $-1.07\pm 0.03$ & $      -      $\\
  IC~5298    & $4.46\pm 0.02$ & $2.53\pm 0.02$ & $3.16\pm 0.02$ & $-0.79\pm 0.02$ & $-0.74\pm 0.05$ & $-0.81\pm 0.04$ & $-0.94\pm 0.12$ & $      -      $ & $-0.76\pm 0.07$ & $      -      $\\
  MRK~331    & $8.71\pm 0.03$ & $2.52\pm 0.01$ & $3.02\pm 0.02$ & $-0.70\pm 0.02$ & $-0.67\pm 0.05$ & $-0.64\pm 0.03$ & $-0.59\pm 0.12$ & $      -      $ & $-0.81\pm 0.05$ & $      -      $\\

\hline
Mean & & $2.441\pm 0.004$ & $2.906\pm 0.005$ & $-0.671\pm 0.007$ & $-0.521\pm 0.018$ & $-0.596\pm 0.007$ & $-0.698\pm 0.040$ & $-0.790\pm 0.079$ & $-0.814\pm 0.020$ & $-0.842\pm 0.136$\\
\hline
\end{tabular}
\label{tab:alpha}
\end{table*}
\end{landscape}

\section{Results}
\label{sec:results}

\subsection{Radio spectral indices}

Before describing the observed radio spectra, we recall the expected, general form of the radio spectrum for a compact starburst, as outlined in \S~\ref{sec:intro}. At frequencies, $\nu\sim 10\;\rm GHz$, a steep synchrotron spectrum dominates, with $S_{\nu} \propto \nu^{\alpha}$, $\alpha \sim -0.8$. At both lower and higher frequencies the spectrum may be expected to flatten due to free-free absorption (low frequencies) and free-free emission (high frequencies).
   
In fig.~\ref{Fig:radio_spectra} we show the radio spectra for all our sources and table ~\ref{tab:alpha} lists the radio spectral indices, $\alpha$, for the sample objects in several frequency ranges. Also shown are the far-infrared-radio flux ratios, $q = \log(FIR/3.75\times 10^{12}\;\rm Hz)/S_{\nu}$, where $S_{\nu}$ is the radio flux in units of $\rm W\,m^{-2}\,Hz^{-1}$ (Condon et al., 1991a). We will use the notation, $q_{\rm 1.4}$, to refer to this quantity calculated at 1.4~GHz.  

The mean value of $\alpha$ between 1.4 and 4.8~GHz  ($\alpha^{1.4}_{4.8}$ hereafter) is -0.521. This steepens steadily towards higher frequencies, with $\alpha^{4.8}_{8.4} = -0.698$, $\alpha^{8.4}_{22.5} = -0.813$, and $\alpha^{15}_{22.5} = -0.842$ (although only 12 objects have values in this last range). This trend can also be seen in the lower-right panel of fig.~\ref{Fig:radio_spectra}. At low frequencies, the flatter spectra are probably due to free-free absorption. However, the fact that the spectra steepen to frequencies as high as 22.5~GHz is unexpected, since these sources should have very significant free-free emitting components, which would tend to flatten the spectra at high frequencies. Some sources even show spectral indices, $\alpha^{8.4}_{22.5}$, that are steeper than what is expected from pure synchrotron emission. The general trend, and these sources in particular, are further discussed in \S~\ref{radiofits}. 

In Fig.~\ref{Fig:alpha} we show how the radio spectral index is related to $q$. There is a clear tendency for higher values of {\bf $q_{\rm 1.4}$} to be associated with radio spectral indices which are flatter, on average, from 1.4 to 22.5~GHz. There is also a clear correlation between $q_{\rm 1.4}$ and $\alpha^{1.4}_{4.8}$, with a much weaker correlation between $q_{\rm 1.4}$ and $\alpha^{4.8}_{8.4}$. On the other hand, $q_{\rm 8.4}$ is not correlated with either $\alpha^{1.4}_{4.8}$ or $\alpha^{4.8}_{8.4}$. This trend would be observed if the 1.4~GHz radio fluxes were reduced by the effects of free-free absorption. In this case, both $q_{\rm 1.4}$ and $\alpha^{1.4}_{4.8}$ are affected.     The 4.8~GHz flux should be much less affected than the 1.4~GHz flux because the free-free optical depth depends on frequency as $\nu^{-2.1}$. The 8.4~GHz flux is free from the effects of absorption, and so no correlation is seen between $q_{\rm 8.4}$ and $\alpha^{1.4}_{4.8}$ or $\alpha^{4.8}_{8.4}$.

At high frequencies we see a correlation between $q_{\rm 8.4}$ and $\alpha^{8.4}_{22.5}$. The correlation seen  in the lower right panel of fig~\ref{Fig:alpha} shows that the fractional contribution of free-free emission is larger in sources with higher values of $q_{\rm 8.4}$. Therefore, despite the fact that the average source shows a radio spectrum which steepens towards higher frequencies, flatter spectral indices tend to occur in sources with higher $q_{\rm 8.4}$. This is expected if the synchrotron emission is powered by supernova explosions because of the delay between the formation of massive stars and the occurrence of the first supernovae.

The trend for higher values of $q$ to be associated with flatter radio spectra over the whole range from 1.4 to 22.5~GHz, {\bf $\alpha^{1.4}_{22.5}$}, will be used as a constraint on the starburst age in the model fitting of paper II.

\subsection{Possible AGN contribution to the radio fluxes}

If there are ULIRGs in our sample that contain an AGN it is possible that this AGN contributes to the radio flux of the source. If the power-law index of the radio emission is different from that of the star formation we may expect to see a correlation between the radio spectral index and the near-infrared colours, which are sensitive to the presence of an AGN. In fig.~\ref{Fig:J-K} we plot the radio spectral index against the near-infrared colour index (J-K). There is no indication of a general correlation, which suggests that most sources do not contain a radio loud AGN. However, we do note that UGC~8058 (Mrk~231) which has the highest value of (J-K) also has a very flat radio spectral index. As mentioned above, there is evidence that the AGN in this source is radio loud. 

\subsection{Radio spectral indices and IRAS colour, $(f_{60}/f_{100})$}

In fig.~\ref{Fig:IRAS_colours} we plot the radio spectral indices against the IRAS colour $(f_{60}/f_{100})$. $\alpha^{1.4}_{4.8}$ shows a correlation with IRAS colour (in good agreement with the results of Sopp \& Alexander, 1992) but this correlation almost disappears if $\alpha^{8.4}_{22.5}$ is considered instead. Because warmer IRAS colours are accompanied by larger FIR fluxes (Young et al. 1989, Soifer et al. 1989) warmer IRAS colours imply higher rates of star formation. Therefore sources with warmer colours should have larger masses of ionized gas. However, the weakness of the correlation between $(f_{60}/f_{100})$ and $\alpha^{8.4}_{22.5}$ suggests that this effect is not the main driver of the stronger correlation seen with $\alpha^{1.4}_{4.8}$. 

The compactness of the regions is probably what causes the correlation in the upper panel of fig.~\ref{Fig:IRAS_colours}. If warmer IRAS colours were associated with more compact sources, then greater free-free absorption would be seen in warmer sources and a correlation between $(f_{60}/f_{100})$ and $\alpha^{1.4}_{4.8}$  would result, even if the mass of ionized gas does not change with $(f_{60}/f_{100})$. Source geometry is considered in more detail in \S~\ref{sec:discussion}.

\begin{figure*}
\centering
\includegraphics[angle=0, width = 18cm]{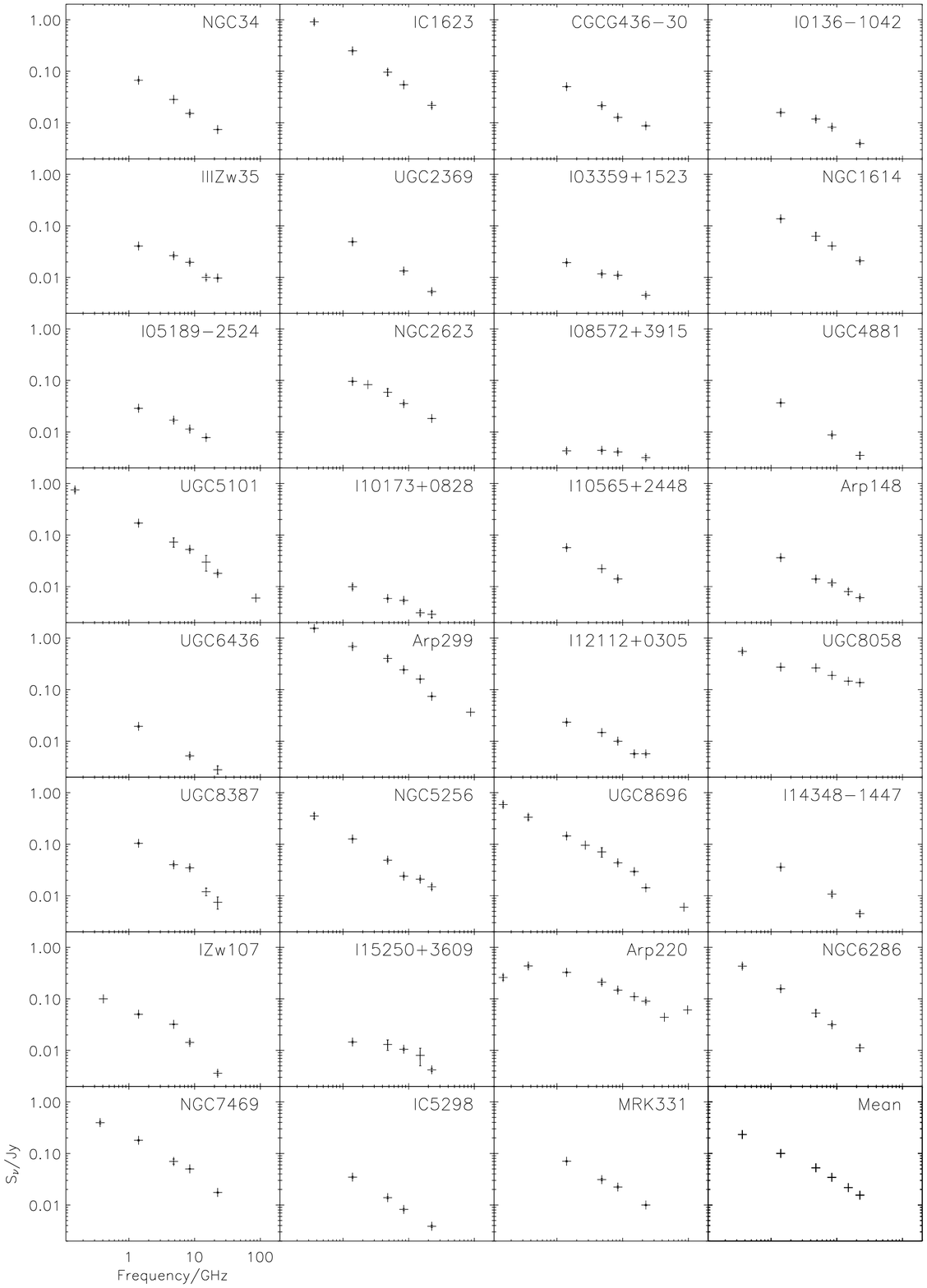}
\caption{The radio spectra of the sample galaxies from 100~MHz to 100~GHz. The lower left panel shows the mean spectrum of all the sources scaled so that the 1.4~GHz flux is $0.1\;\rm Jy$.}
\label{Fig:radio_spectra}
\end{figure*}

\begin{figure*}
\centering
\includegraphics[angle=90, width = 18cm]{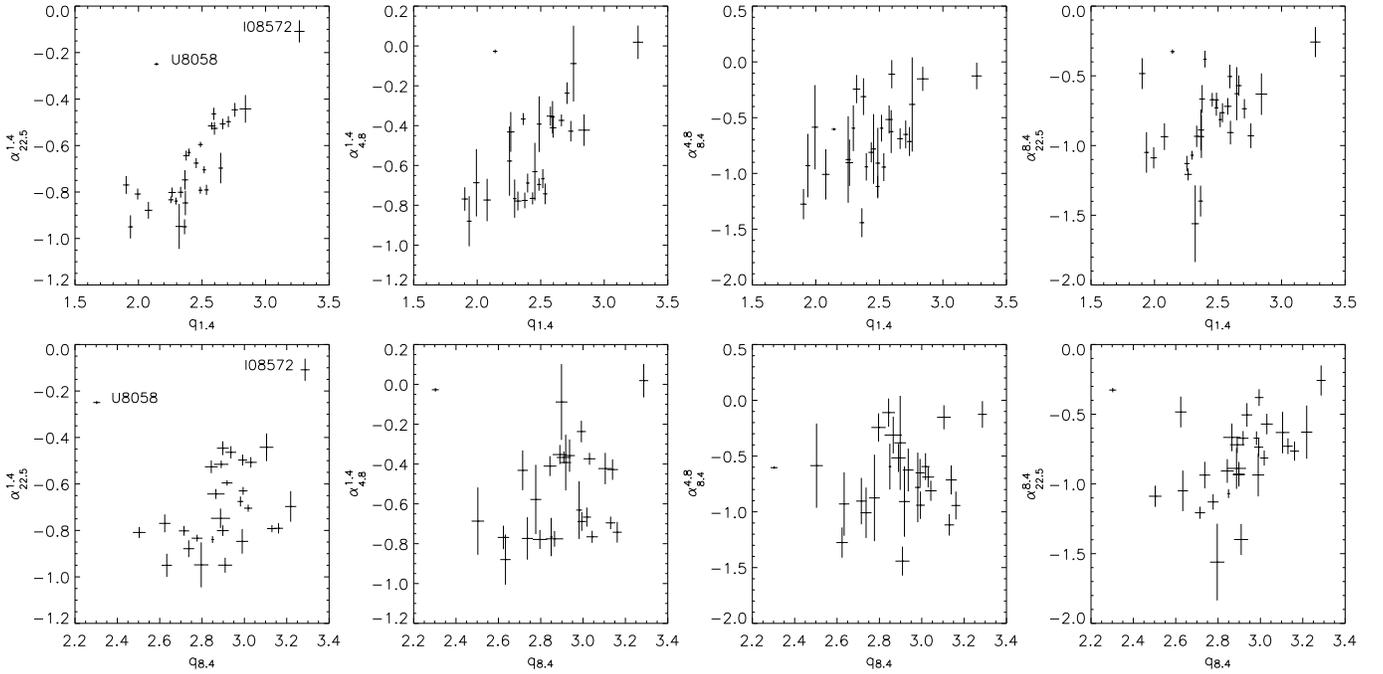}
\caption{Relationship between radio spectral index at various frequencies and the logarithmic FIR-radio flux density ratio q, defined as $q = \log(FIR/3.75\times 10^{12}\;\rm Hz)/S$, where $S$ is the radio flux in units of $\rm W\,m^{-2}\,Hz^{-1}$. The top panels show the radio spectral index as a function of q measured at $1.4\;\rm GHz$ and the bottom panels show the radio spectral indices as a function of q measured at $8.4\;\rm GHz$. The left-most panels show the radio spectral index over the whole range from 1.4 to $22.5\;\rm GHz$ while the remainder show the spectral indices over a more restricted range.}
\label{Fig:alpha}
\end{figure*}

\begin{figure}
\centering
\includegraphics[angle=90, width = 9cm]{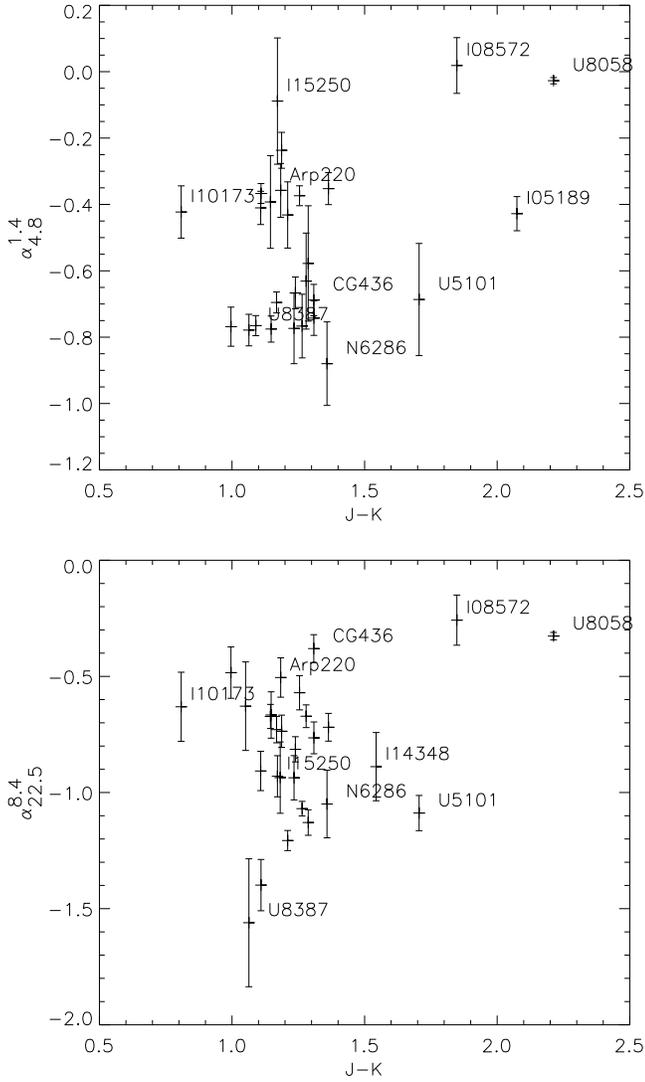}
\caption{The relation between the near-infrared colour J-K and the radio spectral index between 1.4 and 4.8~GHz (top) and 8.4 and 22.5~GHz (bottom). Objects at the extremes of the distributions are labeled.}
\label{Fig:J-K}
\end{figure}

\begin{figure}
\centering
\includegraphics[angle=90, width = 9cm]{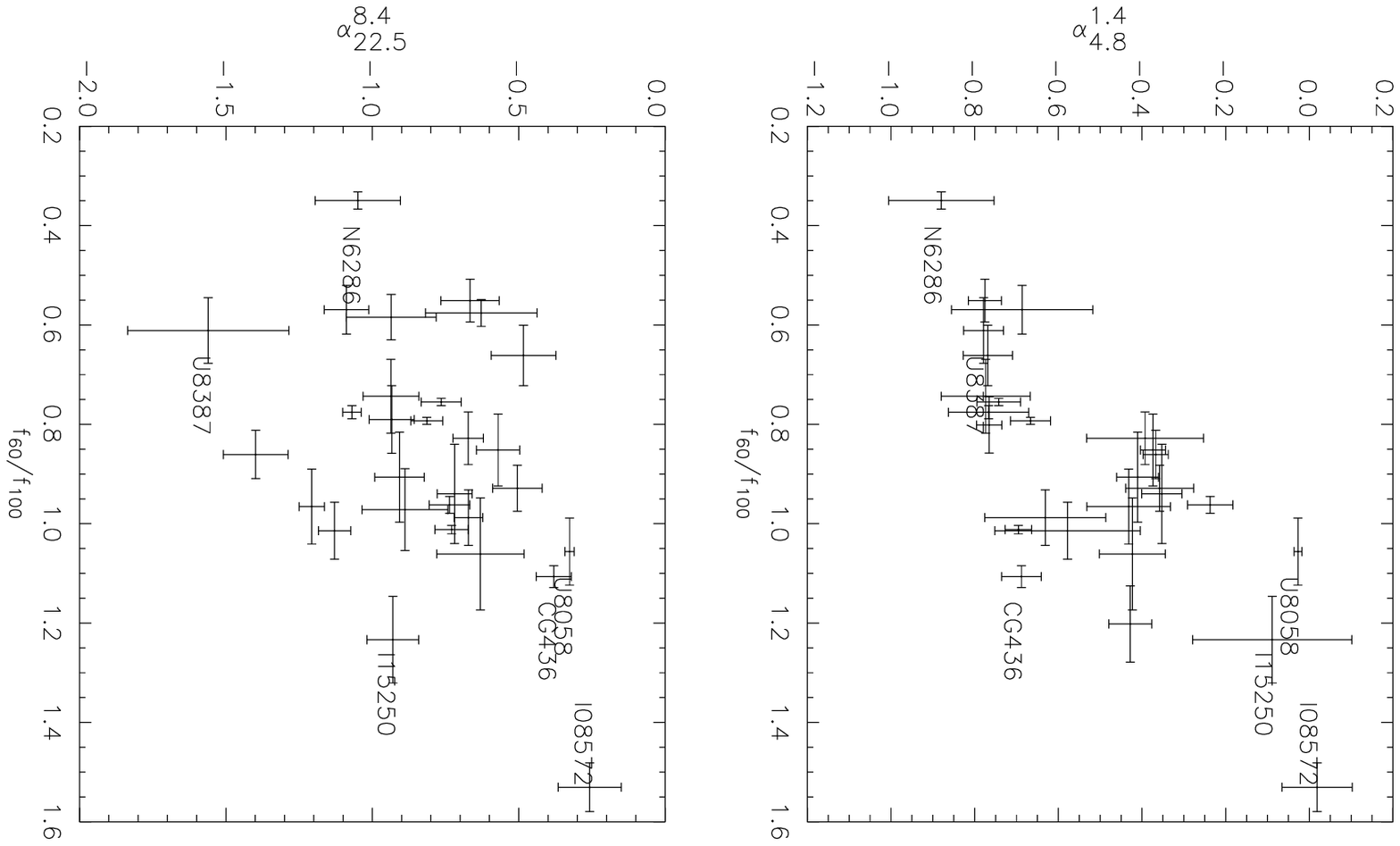}
\caption{{\bf Top:} Correlation between the radio spectral index between 1.4 and 4.8~GHz and IRAS 60/100 micron flux ratio. {\bf Bottom:} Same plot for the 8.4-22.5~GHz spectral index. Objects at the extremes of the distributions are labeled. Note that that UGC~8058 may have a radio loud AGN.}
\label{Fig:IRAS_colours}
\end{figure}

\section{Discussion}
\label{sec:discussion}

\subsection{The radio spectral index and the FIR-radio flux ratio, $q$}

In fig.~\ref{Fig:radio_spectra} there are several sources in which both a low and a high frequency spectral flattening are seen (e.g. IRAS~12112+0305). This flattening is due to free-free absorption at low frequencies and free-free emission at high frequencies. However, there are also sources which show evidence of free-free emission at high frequencies, with no sign of free-free absorption at low frequencies (e.g.CGCG436) and others that show signs of free-free absorption, with no sign of free-free emission (e.g. NGC~2623). The simplest explanation for these contrasting radio spectra is probably variations in the compactness of the emitting regions.               

For a region of radius $r$ and density $N_{H}$ maintained in ionization equilibrium by a source of ionizing radiation of fixed luminosity, the total ionization/recombination rate, $(4/3) \pi r^3N_{H}^2 \alpha_{B}$, where $\alpha_B$ is the recombination coefficient, is constant (Osterbrock, 1989). The free-free optical depth, $\tau_{\rm ff} \propto n_e^2\,r$, for a density of ionized gas $n_e$. Therefore, as long as $\tau_{\rm ff}$ remains below unity (certainly true at 22~GHz) the \emph{integrated} free-free emission, {\bf $\propto \tau_{\rm ff}\,r^2$}, from a region with fixed ionizing luminosity, is independent of the source size. 
        
However, where this gas absorbs a luminosity source, the absorption is proportional to $e^{-\tau_{\rm ff}}$ (foreground screen) and the dependence of $\tau_{\rm ff}$ on source size means that the compactness of a region very strongly affects the amount of absorption. For a given ionizing luminosity and synchrotron luminosity, an extended region may show no evidence of free-free absorption at 1.4~GHz, whereas a more compact region will.

This effect is probably behind the the diverse spectral shapes of fig.~\ref{Fig:radio_spectra}   and is also important in understanding fig.~\ref{Fig:alpha}. 

In considering the relation between the radio spectral indices at various frequencies and $q$, shown in fig.~\ref{Fig:alpha}, it is useful to keep in mind how each are related to the star formation rate. Both the FIR and the integrated free-free emission are expected to be proportional to the current star formation rate. The synchrotron emission is also proportional the star formation rate, but as the emission is delayed by main sequence lifetime of massive stars, it measures the star formation rate $\sim 10^7$ years prior to the current epoch. Therefore, in objects characterized by bursts of star formation, such as those in our sample, we expect both $\alpha$ and $q$ to be a function of the age of the starburst. Values of $q$ are higher where the current star formation rate is greater than that $\sim 10^7$ years ago, and lower where the star formation rate has declined. In this scenario, $q$ is a measure of the ``age'' of a starburst, with higher values associated with younger objects.

In fig.~\ref{Fig:alpha}, as well as the correlations between  $q_{1.4}$ and $\alpha^{1.4}_{4.8}$, and between  $q_{8.4}$ and $\alpha^{8.4}_{22.5}$, we note the \emph{lack} of a correlation between $q_{8.4}$ and $\alpha^{1.4}_{4.8}$. If large values of $q_{8.4}$ are due to a delay in the production of synchrotron emission relative to that of thermal dust and free-free emission,  as the correlation between  $q_{8.4}$ and $\alpha^{8.4}_{22.5}$ suggests, then the lack of correlation between  $q_{8.4}$ and $\alpha^{1.4}_{4.8}$ implies that the age of the starburst is not the only effect.   

Because source compactness strongly affects the amount of free-free absorption, but does not affect the free-free emission (for $\tau_{\rm ff} \ll 1$) the presence of free-free emission at 22.5~GHz does not imply the presence of free-free absorption at 1.4~GHz. Therefore, although the youngest sources (highest $q_{8.4}$ values) have larger ionized gas fractions, there exists a range of source compactness for any given age. This causes the lack of correlation between $q_{8.4}$ and $\alpha^{1.4}_{4.8}$.

\subsection{Low frequency radio spectral index and the FIR colour}
As we saw in fig.~\ref{Fig:IRAS_colours} the low frequency radio spectral index $\alpha^{1.4}_{4.8}$ is correlated with the FIR flux ratio $f_{60}/f_{100}$. Warmer FIR colours are therefore associated with more free-free absorption. However, we recall that \emph{younger} sources do not show more free-free absorption (Fig.~\ref{Fig:alpha}, $\alpha^{1.4}_{4.8}$ vs. $q_{8.4}$). Together, these imply that the warmest dust is not found in the youngest sources, as has been suggested previously (Vega et al., 2005). If source compactness has an important role to play in determining the level of free-free absorption at low radio frequencies, then the \emph{most compact} are those in which the dust is warmest.

\subsection{High frequency radio spectral indices}
\label{radiofits}

A small number of ULIRGs in our sample show surprisingly steep spectral indices at high frequencies.
UGC~5101, Arp~299, UGC~8387, UGC~8696, IZw~107, NGC~6286 and NGC~7469 all have radio spectral indices between 8.4 and 22.5~GHz steeper than $\alpha=-1$. There appears to be either a deficit in the observed flux at 22~GHz or a spectrum that is also very steep at lower frequencies (UGC~5101, IZw~107, NGC~6286). In the case of IZw107 the radio spectrum has a spectral index of $\alpha=-1.4$ from 4.8 to 22.5~GHz\footnote{
Note, however, that a steeper power-law fit to the high frequency radio data would 
not fit the $1.4\;\rm GHz$ and $408\;\rm MHz$ data points}. In those cases where the radio spectra show a steepening towards higher frequencies the spectrum appears to show a break around 15~GHz.

This is contrary to what we expect in sources with a large mass of free-free emitting gas. The radio spectrum should flatten at high frequencies due to the increasing importance of the thermal emission ($S_\nu \propto \nu^{-0.1}$). Although it is conceivable that the mass of free-free emitting gas is reduced by the absorption of ionizing photons by dust, this effect alone cannot produce an observed radio spectrum that steepens towards higher frequencies. 

A steepening of the radio spectrum towards higher frequencies is expected in the
presence of `aging' of the relativistic electron population (Condon, 1992).
Because synchrotron (and inverse Compton) losses are greater at higher
frequencies ($\tau_{\rm sync} \propto \nu^{-0.5}$) higher energy electrons emit
synchrotron photons for shorter times than those of lower energies.  For an
instantaneous injection of relativistic electrons the radio spectrum therefore
steepens with increasing frequency. The `break' frequency which parameterizes
this steepening moves to lower frequencies at later times after the injection.
Spectral aging effects were considered by Colbert, Wilson \& Bland-Hawthorn (1994) 
to explain the steep radio spectral index in the non-nuclear regions of NGC~6240.

In ULIRGs the electron lifetimes are very short because in a high
radiation energy density environment inverse Compton losses would remove all high
energy electrons before they could emit synchrotron photons. The observation of
synchrotron emission in such sources therefore shows that they must have strong
B-fields and short electron lifetimes due to synchrotron losses. Electron
lifetimes in the GHz region are shorter not only than a starburst timescale ($10^7 - 10^8$ yr) but
also than the lifetime of the most massive stars ($\sim 5 \times 10^6\;\rm yr$).
Therefore the production of relativistic electrons cannot be considered
instantaneous, but is rather close to a continuous injection approximation.

Under the assumption of continuous injection of relativistic electrons, the radio 
spectrum can be considered as the superposition of several instantaneously injected 
populations, each with a different age, and therefore each with a different `break' 
frequency. The spectrum which results from the summation of all these components
is straight. If the asymptotic change in spectral index for a single component
of spectral index, $\alpha$, is $\Delta \alpha = -0.5$ (such as in the ``dynamical 
halo'' model of Lerche \& schlickeiser, 1981) then the resulting spectrum has a 
spectral index $\alpha_{\rm aged} = \alpha - 0.5$. The radio spectra of all 
ULIRGs should therefore be steepened by the effects of spectral aging, but we do not 
expect to detect a down-turn in the spectra towards high frequencies if continuous 
injection is a good approximation. We note however, that because the spectral break 
for an aging, instantaneously injected electron population spends little time in the 
$\sim 10\;\rm GHz$ range (where we observe the putative breaks) the form of the 
time-averaged spectrum is rather sensitive to stochastic effects in the recent past. 
Such effects will be important where the supernova rate shows variations on timescales 
of order the electron lifetime.

Qualitatively, such a stochastic effect could be caused by a
radio hypernova, that might have an order of magnitude greater flux 
than a typical supernova. However, it seems very unlikely that a single such 
source could effect the integrated radio spectrum of an entire ULIRG. The compact 
sources mapped in Arp~220 by Lonsdale et al. (2006) account for only 6\% of the total 
flux at $18\;\rm cm$

Lisenfeld et al. (2004) considered various explanations for the high frequency 
turn-over seen in the synchrotron  spectrum of the starburst dwarf galaxy NGC~1569, once 
the contribution of thermal emission had been subtracted. They 
concluded that  a sharp turn-over could only be achieved by either 
a rapid temporal variation of the star formation rate or through the preferential 
escape of low-energy electrons from the galaxy's disk. In NGC~1569 a starburst that 
started abruptly $10^7\;\rm yr$ ago was found to reproduce the observed break
in the synchrotron spectrum.
The escape of low energy electrons from the galaxy can produce a break if convection
can transport the electrons out of the disk during their radiative lifetime. As the lower 
energy electrons suffer less synchrotron losses, and therefore emit for 
longer than high-energy electrons, they can in principle be lost preferentially and cause
a break in the spectrum. Lisenfeld et al. (2004) found that a convective wind velocity of 
$150\;\rm km\,s^{-1}$ can reproduce the spectral break in NGC~1569.  
However, the lifetimes of electrons against radiative losses in ULIRGs should be 
considerably shorter than in NGC~1569. Under the assumption of energy equipartition 
between the radiation and magnetic fields, Condon et al. (1991a) estimate electron lifetimes 
of $\sim 10^4\;\rm yr$ for electrons radiating at $8.4\;\rm GHz$ in the same sample of ULIRGs 
from which ours were selected, while that in NGC~1569 was estimated to be 
$5\times 10^6\;\rm yr$ by Lisenfeld et al. Because the supernova rate cannot change 
significantly over timescales of $10^4\;\rm yr$\footnote{Type 1b and type II supernovae are thought to be the source of the synchrotron emission in ULIRGs, and occur by the core-collapse of stars more massive than $8\;\rm M_{\odot}$. The difference in lifetime between a star of $8\;\rm M_{\odot}$ and a star of $100\;\rm M_{\odot}$ is $\sim 5 \times 10^7\;\rm yr$. Therefore, for a normal initial mass function, even if the stars formed instantaneously, the supernova rate cannot show large variations on timescales as short as $10^4\;\rm yr$.} and a convective wind could not remove
electrons from the sources over similar timescales, none of these mechanisms seem a plausible
explanation for a high frequency spectral break in the present sample of ULIRGs.

Although there is no doubt that some sources have steep spectral indices from 1 to $20\;\rm GHz$, in the majority of cases where a turn-over is seen it is due only to the 22.5~GHz data point (Arp~299, UGC~8696, IRAS~15250+3609). We would require data at a higher frequency (e.g. 43~GHz) for these sources in order to be sure of the spectral bend and warrant a more in-depth discussion.

\section{Conclusions}
\label{sec:conclusions}
We have presented new high frequency radio data for a sample of ULIRGs, that, together with re-reduced archival data and fluxes taken from the literature forms a sample of 31 ULIRGS with well-sampled radio spectra. All but 1 source have a measured flux at 22.5~GHz. Every effort has been made to ensure that the fluxes are reliable measures of the integrated radio emission. This means that the possibility of missing flux in high-resolution interferometer data and of confusing sources in low resolution data has been considered. We find the following,

\begin{itemize}
\item Few sources have straight power-law slopes. Although the individual spectra show a variety of spectral bends the trend of the mean spectrum is a steepening from 1.4~GHz to 22.5~GHz. Above and below these frequencies data is too sparse to define a meaningful average.
\item The low frequency radio spectral index between 1.4 and 4.8~GHz is correlated with  the FIR-radio flux density ratio q calculated at 1.4~GHz. This correlation is due to free-free absorption of low frequency synchrotron photons by ionized gas.
\item The high frequency radio spectral index between 8.4 and 22.5~GHz is correlated with the FIR-radio flux density ratio q, calculated at 8.4~GHz. This correlation is due to a higher fractional contribution from free-free emission in sources with higher values of $q_{8.4}$. These are the younger sources.
\item Given the above correlations the \emph{lack} of a correlation between the radio spectral index between 1.4 and 4.8~GHz and $q_{8.4}$ is evidence that the compactness of the emitting regions plays an important role in defining the slope of the low frequency radio spectrum.  
\item The fact that the low frequency radio spectral index $\alpha^{1.4}_{4.8}$ is correlated with the FIR flux ratio $f_{60}/f_{100}$ (while $\alpha^{8.4}_{22.5}$ is not) suggests that the ULIRGs in which the emitting regions are most compact are those with the warmest IRAS $f_{60}/f_{100}$ colours.

\end{itemize}   
     
If the FIR-to-radio flux ratio measured at 8.4~GHz, $q_{8.4}$, is a measure of the age of the star formation in ULIRGs (larger values being `younger') we find that the youngest objects have a larger contribution to their radio fluxes from free-free emission, but do not have the most free-free absorption. The sources with most free-free absorption are instead those in which the emitting regions are more compact. Although these have the warmest FIR colours, they are not necessarily the youngest sources.

Observations at radio frequencies below 1.4~GHz will be a very effective way of quantifying the effects of free-free absorption in these sources.

In paper 2 these radio spectra, together with data from the radio to the near-infrared, will be used to constrain new models that reproduce the spectral energy distribution of compact starbursts with a possible AGN contribution.

\vspace{2cm}

\noindent

\begin{acknowledgements}
MC acknowledges the  the support of an I.N.A.F. research fellowship.
O.V.  acknowledges the support of the INAOE and the Mexican CONACYT 
projects 36547 E and 39714 F. A.B. acknowledges the warm hospitality of INAOE.
A.B., G.L.G., and L.S. acknowledge partial funding by the European Community
by means of the Maria Curie contract MRTN-CT-2004-503929, `MAGPOP'.
The National Radio
Astronomy Observatory is a facility of the National Science Foundation
operated under cooperative agreement by Associated Universities,
Inc. This publication makes use of data products from the Two Micron
All Sky Survey, which is a joint project of the University of
Massachusetts and the Infrared Processing and Analysis
Center/California Institute of Technology, funded by the National
Aeronautics and Space Administration and the National Science
Foundation. This research has made use of the NASA/IPAC Extragalactic Database (NED) which is operated by the Jet Propulsion Laboratory, California Institute of Technology, under contract with the National Aeronautics and Space Administration.
\end{acknowledgements}


\begin{appendix}

\begin{table}
\label{tab:radiofluxes}
\caption{Radio fluxes.}
\centering
\begin{tabular}{l c c c}
\hline\hline
Source & Frequency & Flux & Ref. \\
   & (GHz) & (mJy) &  \\
\hline
NGC~34		   &  1.4  & $66.9 \pm 2.5$ &	d \\
		   &  4.8  & $28.4 \pm 0.3$ &	c \\
		   &  8.4  & $15.2 \pm 0.8$ &	e \\
		   & 22.5  & $7.41 \pm 0.14$ &	b \\
IC~1623		   & 0.365 & $906 \pm 83$ &	v \\
		   &  1.4  & $249 \pm 9.8$ &	d \\
		   &  4.8  & $96 \pm 12$ &	i \\
		   &  8.4  & $54.6 \pm 1$ &	a \\
		   & 22.5  & $21.7 \pm 2$ &	a \\
CGCG436-30	   &  1.4  & $49.8 \pm 1.5$ &	d \\
		   &  4.8  & $21.5$	 &	g \\
		   &  8.4  & $12.7 \pm 0.6$ &	e \\
		   & 22.5  & $8.73 \pm 0.31$ &	b \\
IR~01364-1042	   &  1.4  & $15.8 \pm 0.7$ &	d \\
		   &  4.8  & $11.8$	 &	o \\
		   &  8.4  & $8.2 \pm 0.4$ &	e \\
		   & 22.5  & $3.97 \pm 0.19$ &	b \\
IIIZw~35	   &  1.4  & $40.6 \pm 1.3$ &	d \\
		   &  4.8  & $26.3$	 &	g \\
		   &  8.4  & $19.7$	 &	e \\
		   &  15   & $10.0 \pm 1$ &	c \\
		   & 22.5  & $9.7 \pm 0.25$ &	b \\
UGC~2369	   &  1.4  & $49 \pm 1.5$ &	d \\
		   &  8.4  & $13.3$	 &	e \\
		   & 22.5  & $5.3 \pm 0.3$ &	a \\
IR~03359+1523$^{+}$&  1.4  & $19.4 \pm 0.7$ &	d \\
		   &  4.8  & $11.7$	 &	g \\
		   &  8.4  & $11.0$	 &	e \\
		   & 22.5  & $4.5 \pm 0.3$ &	a \\
NGC~1614	   &  1.4  & $137.1 \pm 4.9$ &	d \\
		   &  4.8  & $63 \pm 11$ &	j \\
		   &  8.4  & $40.7 \pm 0.3$ &	c \\
		   & 22.5  & $21.0 \pm 1$ &	a \\
IR~05189-2524	   &  1.4  & $28.8 \pm 1.0$ &	d \\
		   &  4.8  & $17.0 \pm 0.9$ &	n \\
		   &  8.4  & $11.4$	 &	e \\
		   &  15   & $7.8 \pm 0.2$ &	c \\
NGC~2623	   &  1.4  & $95.7 \pm 2.9$ &	d \\
		   &  2.4  & $83.0 \pm 4$ &	l \\
		   &  4.8  & $59.0 \pm 10$ &	h \\
		   &  8.4  & $35.5$	 &	e \\
		   & 22.5  & $18.3 \pm 0.2$ &	a \\
IR~08572+3915	   &  1.4  & $4.3 \pm 0.4$ &	d \\
		   &  4.8  & $4.4 \pm 0.2$ &	n \\
		   &  8.4  & $4.1 \pm 0.2$ &	e \\
		   & 22.5  & $3.18 \pm 0.30$ &	b \\
UGC~4881	   &  1.4  & $36.8 \pm 1.2$ &	d \\
		   &  8.4  & $8.8$	 &	e \\
		   & 22.5  & $3.5 \pm 0.5$ &	a \\
\hline
\end{tabular}

$^+$ Only one source of a close pair emits in the radio.
$^*$ Source possibly variable in radio. 1.4, 4.8, 8.4, 15 and 22.5~GHz fluxes
    all made on the same date.
References.
{\bf (a)} this work.
{\bf (b}) Prouton et al. (2004).
{\bf (c)} reduced from VLA archive. See table~\ref{tab:archive} for details.
{\bf (d)} NVSS, Condon et al. (1998).
{\bf (e)} Condon et al. (1991a).
{\bf (f)} Sopp \& Alexander (1991).
{\bf (g)} Sopp \& Alexander (1992).
{\bf (h)} Gregory \& Condon (1991).
{\bf (i)} Griffith et al. (1994).
{\bf (j)} Griffith et al.(1995).
{\bf (k)} Condon et al.(1983).
{\bf (l)} Dressel \& Condon (1978).
{\bf (m)} Condon, Frayer \& Broderick (1991b) .
{\bf (n)} Rush, Malkan \& Edelson (1996).
{\bf (o)} Baan \& Kl{\"o}ckner (2006).
{\bf (p)} Crawford et al. (1996).
{\bf (q)} Sramek \& Tovmassian (1976).
{\bf (r)} Imanishi, Nakanishi \& Kohno (2006).
{\bf (s)} Zhao et al. (1996).
{\bf (t)} Rodriguez-Rico et al. (2005).
{\bf (u)} Anantharamaiah et al. (2000). 
{\bf (v)} Douglas et al. (1996)
{\bf (w)} Ficarra, Grueff \& Tomassetti (1985)
\end{table}

\begin{table}
\addtocounter{table}{-1}
\caption{cont.d.}
\centering
\begin{tabular}{l c c c}
\hline\hline
Source & Frequency & Flux & Ref. \\
   & GHz & mJy &  \\
\hline
UGC~5101	   & 0.151 & $750 \pm 60$ &	f \\
		   &  1.4  & $170.1 \pm 5.8$ &	d \\
		   &  4.8  & $73 \pm 15$ &	f \\
		   &  8.4  & $52.6$	 &	e \\
		   &  15   & $30 \pm 10$ &	f \\
		   & 22.5  & $18.0 \pm 1$ &	a \\
		   & 85.5  & $6$	 &	r \\
IR~10173+0828	   &  1.4  & $9.9 \pm 0.9$ &	d \\
		   &  4.8  & $5.88 \pm 0.2$ &	c \\
		   &  8.4  & $5.4$	 &	e \\
		   &  15   & $3.1 \pm 0.3$ &	c \\
		   & 22.5  & $2.9 \pm 0.4$ &	a \\
IR~10565+2448	   &  1.4  & $57.0 \pm 2.1$ &	d \\
		   &  4.8  & $22.21 \pm 0.13$ &	p \\
		   &  8.4  & $14.1$	 &	e \\
Arp~148		   &  1.4  & $36.4 \pm 1.2$ &	d \\
		   &  4.8  & $14.0 \pm 0.5$ &	c \\
		   &  8.4  & $11.76 \pm 1$ &	a \\
		   &  15   & $8.0 \pm 1$ &	c \\
		   & 22.5  & $6.1 \pm 0.3$ &	a \\
UGC~6436	   &  1.4  & $19.4 \pm 0.7$ &	d \\
		   &  8.4  & $5.2 \pm 0.3$ &	a \\
		   & 22.5  & $2.8 \pm 0.5$ &	a \\
Arp~299		   & 0.365 & $1550 \pm 50$   &    v \\
		   &  1.4  & $686.3 \pm 25.4$ &	d \\
		   &  4.8  & $403 \pm 45$ &	h \\
		   &  8.4  & $243 \pm 8$ &	c \\
		   &  15   & $160 \pm 3$ &	c \\
		   & 22.5  & $74 \pm 2$	 &	c \\
IR~12112+0305	   &  1.4  & $23.3 \pm 0.8$ &	d \\
		   &  4.8  & $14.7 \pm 0.2$ &	c \\
		   &  8.4  & $10.0$	 &	e \\
		   &  15   & $5.7 \pm 0.2$ &	c \\
		   & 22.5  & $5.7 \pm 0.3$ &	a \\
UGC~08058$^{*}$	   & 0.365 & $551 \pm 39$ &	v \\
		   &  1.4  & $274 \pm 3$  &	c \\
		   &  4.8  & $265 \pm 1$ &	c \\
		   &  8.4  & $189 \pm 1$ &	c \\
		   &  15   & $146 \pm 1$ &	c \\
		   & 22.5  & $137 \pm 2$ &	c \\
UGC~8387	   &  1.4  & $104.4 \pm 3.2$ &	d \\
		   &  4.8  & $46$	 &	m \\
		   &  8.4  & $34.9$	 &	e \\
		   &  15   & $12 \pm 2$	 &	c \\
		   & 22.5  & $7.5 \pm 2$ &	a \\
NGC~5256	   & 0.365 & $353 \pm 43$ &	v \\
		   &  1.4  & $126.3 \pm 4.5$ &	d \\
		   &  4.8  & $43.3 \pm 3.1$ &	n \\
		   &  8.4  & $24.1 \pm 1$ &	c \\
		   &  15   & $21.0 \pm 1$ &	c \\
		   & 20.0  & $14.9 \pm 1.5$ &	n \\
UGC~8696	   & 0.151 & $590 \pm 60$ &	f \\
		   & 0.365 & $335 \pm 37$ &	v \\
		   &  1.4  & $144.7 \pm 5.1$ &	d \\
		   &  2.7  & $96 \pm 6$	 &	q \\
		   &  4.8  & $71 \pm 15$ &	f \\
		   &  8.4  & $43.5$	 &	e \\
		   &  15   & $29.5 \pm 1$ &	c \\
		   & 22.5  & $14.3 \pm 0.3$ &	a \\
		   & 85.6  & $6$	 &	r \\
IR~14348-1447	   &  1.4  & $35.9 \pm 1.2$ &	d \\
		   &  8.4  & $10.8 \pm 1$ &	a \\
		   & 22.5  & $4.5 \pm 0.5$ &	a \\
\hline
\end{tabular}
\end{table}

\begin{table}
\addtocounter{table}{-1}
\caption{cont.d.}
\centering
\begin{tabular}{l c c c}
\hline\hline
Source & Frequency & Flux & Ref. \\
   & GHz & mJy &  \\
\hline
IZw~107		   & 0.408 & $100 \pm 20$ &	w \\
		   &  1.4  & $50.3 \pm 1.6$ &	d \\
		   &  4.8  & $32.0 \pm 0.6$ &	h \\
		   &  8.4  & $14.3 \pm 1$ &	a \\
		   & 22.5  & $3.6 \pm 0.3$ &	a \\
IR~15250+3609	   & 0.151 & $<100$	 &	f \\
		   &  1.4  & $14.5 \pm 0.6$ &	d \\
		   &  4.8  & $13 \pm 3$	 &	f \\
		   &  8.4  & $10.5$	 &	e \\
		   &  15   & $8 \pm 3$	 &	f \\
		   & 22.5  & $4.2 \pm 0.3$ &	a \\
Arp~220		   & 0.151 & $260 \pm 30$ &	f \\
		   & 0.365 & $435 \pm 26$ &	v \\
		   &  1.4  & $326.3 \pm 9.8$ &	d \\
		   &  2.4  & $312 \pm 16$ &	l \\
		   &  2.7  & $260 \pm 13$ &	k \\
		   &  4.8  & $210 \pm 20$ &	f \\
		   &  8.4  & $148.0$	 &	e \\
		   &  15   & $110 \pm 1$ &	f \\
		   & 22.5  & $90 \pm 6$	 &	s \\
		   &  43   & $44 \pm 4$	 &	t \\
		   & 97.2  & $61 \pm 10$ &	u \\
NGC~6286	   & 0.365 & $431\pm 37$ &	v \\
		   &  1.4  & $156.7 \pm 5.6$ &	d \\
		   &  4.8  & $53.0 \pm 8$ &	h \\
		   &  8.4  & $31.5 \pm 1.6$ &	a \\
		   & 22.5  & $11.2 \pm 1.5$  &	a \\
NGC~7469	   & 0.365 & $397 \pm 24$ &	v \\
		   &  1.4  & $180.5 \pm 5.4$ &	d \\
		   &  4.8  & $70 \pm 8$	 &	g \\
		   &  8.4  & $50.2 \pm 0.6$ &	c \\
		   & 22.5  & $17.50 \pm 0.5$ &	b \\
IC~5298		   &  1.4  & $34.7 \pm 1.4$ &	d \\
		   &  4.8  & $13.9$	 &	g \\
		   &  8.4  & $8.2 \pm 0.4$ &	e \\
		   & 22.5  & $3.86 \pm 0.18$ &	b \\
Mrk~331		   &  1.4  & $70.7 \pm 2.2$ &	d \\
		   &  4.8  & $31.1$	 &	g \\
		   &  8.4  & $22.3 \pm 1$ &	a \\
		   & 22.5  & $10.0 \pm 0.3$ &	b \\
\hline
\end{tabular}
\end{table}

\section{Individual objects}
Here we briefly describe aspects of the radio emission relevant to the derivation of 
integrated fluxes.
\smallskip

\noindent
{\bf IC~1623}\\
The 22.5~GHz emission shows an unresolved peak at the position of the eastern nucleus
with a clear extension towards the west. The distribution of emission is very similar to
that of the molecular gas (Iono et al. 2004, Yun et al. 1994).
\smallskip

\noindent
{\bf IIIZw~35}\\
The radio emission originates from the more northerly of the 2 optical nuclei.
\smallskip

\noindent
{\bf IRAS~03359+1523}\\
$1\, \farcm 5$ to the south of this object there is a $15.4\; \rm mJy$ radio source.
IRAS~03359+1523 itself emits only $19.4\; \rm mJy$ at 1.4~GHz. Rather than reducing the IRAS fluxes by nearly
 a factor of two we choose to exclude this object from the analysis of paper 2.
\smallskip

\noindent
{\bf UGC~4881}\\
We detect two sources in this object at the same locations as those revealed in the 8.4~GHz
observations of Condon et al. (1991a). The more easterly component is the brighter and accounts
for $\sim 70\%$ of the total flux detected, which is similar to the flux ratio at 8.4~GHz.
\smallskip

\noindent
{\bf UGC~5101}\\
A $12.3\; \rm mJy$ radio source lies $1^{\prime}$ to the north-east. IRAS fluxes reduced by 8\%.
\smallskip

\noindent
{\bf IRAS~10173+0828}\\ 
The sub-arcsecond 15 and 22.5~GHz observations of Smith,
Lonsdale \& Lonsdale (1998) show an unresolved nuclear source. 
\smallskip

\noindent
{\bf UGC~6436}\\ 
The NVSS shows that a companion galaxy, IC~2810b,
located $1\, \farcm 2$ to the south-east has a flux of $7.7\; \rm mJy$
compared to the $19.4\; \rm mJy$ of UGC~6436. This would be confused at
IRAS resolutions and so the IRAS fluxes have been reduced by 26\%. No
other observations were at so low a resolution as to make
contamination from this source a problem.
\smallskip

\noindent
{\bf Arp~299}\\
Actually a close interacting pair consisting of NGC~3690 and IC~694.
\smallskip

\noindent
{\bf IRAS~12112+0305}\\ 
A source $2\, \farcm 4$ to the south-east has a 1.4~GHz 
flux of $5.1\; \rm mJy$ compared to the $23.3\; \rm mJy$ emitted by
IRAS~12112+0305. IRAS fluxes reduced by 15\%. 
\smallskip

\noindent
{\bf UGC~8387}\\
Our 22.5~GHz map shows an unresolved source but the higher resolution 8.4~GHz radio
continuum map of Condon et al. (1991a) shows an elongated structure with an extent
of $\sim 4^{\prime\prime}$. Clemens \& Alexander (2004) find that
free-free absorption flattens the radio spectral index towards the centre of the
source showing there to be a large region of dense ionized gas. 

The 22.5~GHz flux for this source may be incorrect. The calibrator flux for this 
source was only about half that reported in the VLA calibrator database.
\smallskip

\noindent
{\bf IZw~107}\\
Our 8.4~GHz map shows that the extended emission seen at 22.5~GHz is due
to the presence of 2 sources separated by $8^{\prime\prime}$.
\smallskip

\noindent
{\bf Arp~220}\\
Smith et al. (1998) resolved the compact radio emission into separate
supernova remnants and subsequent, high sensitivity VLBI monitoring of
both nuclei has now succeeded in obtaining a direct estimate of the
supernova rate (Lonsdale et al, 2006) of $4 \pm 2\;\rm yr^{-1}$. The
implied star formation rate is sufficient to supply the bolometric
luminosity of the system.
\smallskip

\noindent
{\bf NGC~6286}\\ 
The radio structure at 22 and 8~GHz is elongated in
the same sense as the optical, edge-on disk. A weak source (NGC~6285)
is present $1\, \farcm 5$ to the north-west and this is detected in
our 8.4~GHz data but is undetected in our 22.5~GHz map. The IRAS fluxes
were reduced by 6\% to remove the probable contribution of NGC~6285.

\smallskip

\noindent
{\bf Mrk~331}\\
See Prouton et al. (2004). 

\noindent

\end{appendix}

\end{document}